\documentclass[aps,final,letterpaper,10pt,twocolumn,floats,showpacs,amsmath,amsfonts,amssymb]{revtex4-1}

\usepackage{graphicx}% Include figure files
\usepackage[colorlinks=true,citecolor=blue]{hyperref}
\usepackage[caption=false]{subfig}
\usepackage{pstricks}
\begin{document}
	
	\preprint{APS/123-QED}
	
	\title{Nonrenewal statistics in transport through quantum dots}% Force line breaks with \\
	
	\author{Krzysztof Ptaszy\'{n}ski}
	\email{krzysztof.ptaszynski@ifmpan.poznan.pl}
	\affiliation{%
		Institute of Molecular Physics, Polish Academy of Sciences, ul. M. Smoluchowskiego 17, 60-179 Pozna\'{n}, Poland
	}%
	
	\date{\today}% It is always \today, today,
	
\begin{abstract}
The distribution of waiting times between successive tunneling events is an already established method to characterize current fluctuations in mesoscopic systems. Here, I investigate mechanisms generating correlations between subsequent waiting times in two model systems, a pair of capacitively coupled quantum dots and a single-level dot attached to spin-polarized leads. Waiting time correlations are shown to give an insight into the internal dynamics of the system, for example they allow distinction between different mechanisms of the noise enhancement. Moreover, the presence of correlations breaks the validity of the renewal theory. This increases the number of independent cumulants of current fluctuation statistics, thus providing additional sources of information about the transport mechanism. I also propose a method for inferring the presence of waiting time correlations based on low-order current correlation functions. This method gives a way to extend the analysis of nonrenewal current fluctuations to the systems for which single-electron counting is not experimentally feasible. The experimental relevance of the findings is also discussed, for example reanalysis of previous results concerning transport in quantum dots is suggested. 
		
\end{abstract}
	
	%72.70.+m	Noise processes and phenomena
	%73.63.Kv	Transport in Quantum dots
	%73.23.Hk	Coulomb blockade; single-electron tunneling
	
	\maketitle
	
	%\tableofcontents
	
\section{\label{sec:intro}Introduction}
	
Investigation of current fluctuations is a valuable tool for characterization of the underlying transport mechanism. At low temperatures the main contribution to low-frequency fluctuations comes from the shot noise associated with charge quantization~\cite{blanter2000, nazarov2009}. The magnitude of the shot noise depends on the quasiparticle charge and the correlation of transport events, which is associated with quasiparticle statistics or the presence of interactions~\cite{blanter2000}. In quantum dots the noise was found to provide information about the asymmetry~\cite{gustavsson2006} and spin-dependence~\cite{bulka2000} of tunneling rates in Coulomb blockade systems, interactions with the external environment (with charge~\cite{schaller2010, singh2016}, spin~\cite{sothmann2010} and vibrational~\cite{dong2005} degrees of freedom), the presence of quantum coherence~\cite{kiesslich2007} or many-body quantum correlations (e.g. in the Kondo effect~\cite{delattre2009, ferrier2016}), or the existence of Majorana fermions~\cite{cao2012}. 
	
There are three characteristic regimes of noise distinguished by the value of the Fano factor $F$, which is a ratio of the noise power and the mean current (see e.g. Refs.~\cite{blanter2000, urban2009, emary2012}). If transport events are uncorrelated the noise is Poissonian ($F=1$). In fermionic systems, correlations typically reduce the noise to sub-Poissonian values ($F<1$). However, there are also cases for which the noise may become super-Poissonian ($F>1$).
	
One of the mechanisms generating the super-Poissonian noise is the dynamical channel blockade~\cite{bulka2000, belzig2005}, which is associated with the coexistence of transport channels with different conductances. The electron occupying one channel may block transport through the remaining channels, which would lead to the noise enhancement. This mechanism has been reported in experimental studies of the electronic transport through multilevel quantum dots~\cite{gustavsson2006b, ubbelohde2013, hasler2015} and organic metallic tunnel junctions~\cite{cascales2014}. 
	
Another mechanism of noise enhancement, which should be carefully distinguished from the former, is the stochastic switching between transport channels with different conductances, referred to as the telegraphic switching~\cite{kaasbjerg2015}. It can be caused either by some external factor (e.g. a charge~\cite{schaller2010, singh2016} or spin~\cite{sothmann2010} dynamics of a coupled external system) or by some intrinsic causes (i.e. it can be a result of switching between charge~\cite{fricke2007}, spin~\cite{urban2009}, vibrational~\cite{flindt2005, koch2005, lau2016}, motional~\cite{nishigushi2002} or energy~\cite{kaasbjerg2015} states of a system itself). While this phenomenon has been studied mostly theoretically, the noise enhancement resulting from the switching between the ground and the metastable excited state of a quantum dot~\cite{fricke2007} or from the dependence of tunneling rates on a vibrational state of a single-molecule transistor~\cite{lau2016} has been experimentally observed using charge counting techniques and conventional current measurements, respectively.  
	
A more general analysis of the current fluctuations can be provided by two other already established approaches: the full counting statistics (FCS)~\cite{levitov1993, bagrets2003} and the waiting time distribution (WTD)~\cite{brandes2008}. FCS analyses the distribution of the number of particles transferred in a given time interval, providing information not only about the noise, but also about higher zero-frequency current correlators. High-order cumulants were found to be useful for reconstructing generators of the time evolution of systems with multiple degrees of freedom~\cite{bruderer2014}. In a complementary way, WTD studies the distribution of time delays between subsequent physical events. WTD was found to reveal the short-time dynamics of the system, for example by showing the oscillatory behavior of the distribution associated with a spin precession~\cite{sothmann2014} or coherent oscillations of electrons in a system (e.g. in a quantum dot molecule~\cite{brandes2008} or a quantum dot attached to normal and superconducting leads~\cite{rajabi2013}). In quantum dots both FCS and WTD can be experimentally studied using charge detection techniques~\cite{gustavsson2009}. While such experiments are mostly conducted in ultra-low temperatures (down to several millikelvins), recently the real-time charge trap dynamics monitoring at room temperature has been shown to be feasible~\cite{zbydniewska2015}. However, single electron counting experiments are currently confined to very low currents, with frequencies of tunneling events on the order of kHz~\cite{gustavsson2006, gustavsson2009, ubbelohde2012, haack2015}. Due to this fact, the method determining indirectly WTD with the easier measurable quantities, like the second-order current correlation function, has been proposed~\cite{haack2015}. While this paper focuses on electronic transport, both FCS and WTD are widely applied also beyond this field, for example in quantum optics~\cite{carmichael1989, matthiesen2014, kiilerich2014} and statistical kinetics of biomolecular systems~\cite{chemla2008, moffitt2010, barato2015}.
	
In many cases transport statistics can be well described by the renewal theory, which assumes that successive waiting times between transport events are statistically independent equally distributed random variables~\cite{albert2011}. In such a case, the joint probability density of two successive waiting times $w(\tau_1,\tau_2)$ can be factorized into a product of two single-time distributions $w(\tau_1)w(\tau_2)$~\cite{ dasenbrook2015, budini2010}. When this assumption is satisfied, there are exact identities between cumulants of FCS and WTD, enabling the reconstruction of FCS on the basis of WTD~\cite{albert2011, budini2011}. Moreover, for the renewal dynamics a relation holds between WTD and the second-order correlation function $g^{(2)}(\tau)$, which enables the reconstruction of the former on the basis of the latter, even when the detector efficiency is below 100\%~\cite{carmichael1989}. However, studies of coherent transport in a quantum point contact have shown that waiting times between successive transport events can be correlated~\cite{albert2012, dasenbrook2015}. Correlations were found to appear due to the fermionic statistics of electrons. The presence of correlations indicates that the renewal assumption is not satisfied. In such a case, the aforementioned relations between cumulants of FCS and WTD are no longer valid, as shown by Albert~\textit{et al.}~\cite{albert2012}. Beyond the field of electronic transport, nonrenewal statistics have been previously investigated in some systems described by the Markovian master equation, for example in optical~\cite{cao2006, caycedo-soler2008, budini2010, osadko2011} or biochemical~\cite{saha2011} ones. In particular, the analysis of waiting times correlations was shown to give an insight into the internal dynamics of fluorescent molecules~\cite{cao2006, caycedo-soler2008, osadko2011}. This suggests, that the investigation of physical quantities characterizing the deviation from the renewal behavior may be a useful tool for characterization of the experimentally relevant systems. 
	
The aim of this paper is to extend the investigation of nonrenewal statistics to electronic transport in the sequential tunneling regime. To this end, I analyze two conceptually simple and experimentally feasible transport models: the system of two capacitively interacting quantum dots and the single-impurity Anderson model in the infinite-bias limit realized in a quantum dot attached to spin polarized leads. These models enable a relatively simple qualitative understanding of physical mechanisms generating nonrenewal current fluctuations, and thus may provide a working basis for the analysis of more complex mesoscopic systems. Transport statistics of the studied models are analyzed using the methods based on the master equation~\cite{bagrets2003, brandes2008}. The nonrenewal behavior is made evident by calculating the joint distribution and the cross-correlation of two subsequent waiting times. These quantities are shown to give an insight into the internal dynamics of the system, for example by providing a way to distinguish between the telegraphic switching and the dynamical channel blockade. The influence of the nonrenewal dynamics on FCS and WTD is also analyzed. This reveals some nontrivial effects, like the nonequivalence of waiting time distributions for incoming and outgoing electrons which arises due to the correlation between electron jumps. Moreover, in order to extend the analysis to the systems for which the single-electron counting is not experimentally feasible, I study the influence of the nonrenewal dynamics on the second-order current correlation function and its relation to FCS and WTD. 
	
The paper is organized as follows. Sec.~\ref{sec:methods} describes the methods used to characterize the current fluctuations. In Sec.~\ref{sec:twodots} a model of the double quantum dot system is specified and corresponding results are presented and discussed. Sec.~\ref{sec:anderson} in a similar way deals with transport within the Anderson model. Finally, Sec.~\ref{sec:conclusions} brings conclusions following from my results.
	
%%%%%%%%%%%%%%%%%%%%%%%%%%%%%%%%%%%%%%%%%%%%%%%%%%%%%%%%%%%%%%%%%%%%%%%%%%%%%%%%%%%%%%%%%%%%%%%%%%%%%%%%%%%%%%%
%%%%%%%%%%%%%%%%%%%%%%%%%%%%%%%%%%%%%%%%%%%%%%%%%%%%%%%%%%%%%%%%%%%%%%%%%%%%%%%%%%%%%%%%%%%%%%%%%%%%%%%%%%%%%%%
	
\section{\label{sec:methods}Methods}
The paper reports investigation of electron transport through quantum dot systems in the sequential tunneling regime. I focus on the situations when the coupling to the leads is weak and the separation between the discrete energy levels of the dot and the electrochemical potentials of the leads is high in comparison with $k_B T$ (the assumption referred to as the infinite-bias limit~\cite{brandes2008, emary2012}). As shown in Refs.~\cite{gurvitz1996,gurvitz1998, bagrets2003}, in such a case the tunneling is unidirectional and can be exactly described by the Markovian master equation. The master equation written in the Liouville space takes the form~\cite{carmichael1993, breuer2002}
\begin{equation} \label{mastereq}
\dot{\rho}(t)=\mathcal{L} \rho(t),
\end{equation}
where: $\rho(t)$ is the column vector containing in general both diagonal and non-diagonal elements of the density matrix of the system (state probabilities and coherences), and $\mathcal{L}$ is the square matrix representing the Liouvillian. In all studied systems the coherences are neglected and only the dynamics of diagonal elements of the density matrix is considered. The study is also confined to the systems having a unique stationary state $\rho_0$, which is a solution of the equation $\mathcal{L} \rho=0$.
	
To characterize the current fluctuations in the stationary state I calculate the following quantities: cumulants of the zero-frequency full counting statistics, the waiting time distribution, the joint distribution of two successive waiting times and the second-order current correlation function. To calculate FCS a counting field $\chi$, which counts the number of transport events, is introduced~\cite{levitov1993, bagrets2003}. The master equation is then written in the counting field dependent form
\begin{equation} \label{mastereqchdep}
\dot{\rho}(t,\chi)= \left(\mathcal{L}_0 + \mathcal{J} e^{\chi}  \right) \rho(t, \chi),
\end{equation}
where $\mathcal{J}$ is the operator describing the considered set of jump processes (e.g. tunneling of electrons between a dot and a chosen lead) and the operator $\mathcal{L}_0=\mathcal{L}-\mathcal{J}$ describes the remainder of the system's dynamics. In this paper I focus on time-independent scaled cumulants of FCS. The $n$-th order scaled cumulant $c_n$ is defined in the following way: $c_n = \lim_{t \to \infty} C_n(t)/t$, where $C_n(t)$ is the $n$-th order cumulant of the number of jump events occurring in the time $t$. The scaled cumulants can be calculated as follows~\cite{bruderer2014, wachtel2015}:
\begin{equation} \label{kumkal}
c_n=\left[ \frac{d^n}{d\chi^n} \lambda(\chi)\right]_{\chi = 0},
\end{equation}
where $\lambda(\chi)$ is the scaled cumulant generating function. When the system has the unique stationary state $\rho_0$, $\lambda(\chi)$ is a dominant eigenvalue of the counting-field dependent operator $\mathcal{L}_0 + \mathcal{J} e^{\chi}$~\cite{touchette2009}. In general it is impossible to find analytical expressions for $\lambda(\chi)$ if the rank of the matrix is higher than 4. However, it is not necessary, because all scaled cumulants up to the $N$-th order can be calculated using Eq.~\eqref{kumkal} and solving the following set of linear equations~\cite{bruderer2014, wachtel2015}:
\begin{equation} \label{kumkal2}
\left\{ \frac{d^n}{d \chi^n} \det \left[ \lambda(\chi)-\mathcal{L}_0-\mathcal{J} e^{\chi} \right] \right\}_{\chi = 0}=0,
\end{equation}
with $n$ ranging from 1 to $N$. The knowledge of $\lambda(\chi)$ is not required since $\lambda(0)=0$.
	
Now, the waiting time distribution is considered following the approach developed by T. Brandes~\cite{brandes2008}. One can calculate the distribution of waiting times either between two successive jump events of the same type $k$ or between jumps of two different types $k$ and $l$. In this paper the term "jump of type $k$" refers to the tunneling between a dot and a chosen lead; in general it may refer to an arbitrarily chosen set of transitions within the Markovian model. The Laplace transform of the distribution of waiting times between jumps of types $k$ and $l$ is given by the expression~\cite{brandes2008}
\begin{equation} \label{wkldis}
w_{kl}(s)=\int_0^{\infty} e^{-s \tau} w_{kl}(\tau) d\tau=\frac{\text{Tr}[\mathcal{J}_l (s-\mathcal{L}_0)^{-1} \mathcal{J}_k \rho_0]}{\text{Tr}[\mathcal{J}_k \rho_0]},
\end{equation}
where $\mathcal{L}_0=\mathcal{L}-\mathcal{J}_k$ for $k=l$ and $\mathcal{L}_0=\mathcal{L}-\mathcal{J}_k-\mathcal{J}_l$ for $k \neq l$, with $\mathcal{J}_k$, $\mathcal{J}_l$ being operators describing jumps of types $k$ and $l$ respectively, and $\rho_0$ is the vector of the stationary state. The waiting time distribution $w_{kl}(\tau)$ can be easily obtained from $w_{kl}(s)$ using the inverse Laplace transform. In general $w_{kl}(\tau) \neq w_{lk}(\tau)$. The $n$-th order moments and cumulants of the distribution $w_{kl}(\tau)$, denoted as $\langle \tau_{kl}^n \rangle$ and $\kappa_n^{kl}$ respectively, are given by expressions~\cite{brandes2008}
\begin{eqnarray} \label{momentst}
\langle \tau_{kl}^n \rangle &=& (-1)^n \left[\frac{d^n w_{kl}(s)}{ds^n} \right]_{s=0}, \\ \label{cumst}
\kappa_n^{kl} &=& (-1)^n \left \{\frac{d^n \log[w_{kl}(s)]}{ds^n} \right \}_{s=0}.
\end{eqnarray}
	
In a similar way one can consider a joint distribution of two subsequent waiting times $\tau_{kl}$ and $\tau_{lm}$ separating three successive jumps of types $k$, $l$ and $m$. This distribution is denoted as $w_{klm} (\tau_{kl}, \tau_{lm})$. Distributions of this type for systems described by the master equation have been already considered in Refs.~\cite{caycedo-soler2008, budini2010}. The Laplace transform of the distribution $w_{klm} (\tau_{kl}, \tau_{lm})$ is given by the expression
\begin{eqnarray}
\nonumber w_{klm}(s,z) &=& \int_0^\infty d\tau_{lm} \int_0^\infty d\tau_{kl} e^{-s \tau_{kl}-z \tau_{lm}} w_{klm}(\tau_{kl},\tau_{lm}) \\ 
&=& \frac{\text{Tr}[\mathcal{J}_m (z-\mathcal{L}_0^{lm})^{-1} \mathcal{J}_l (s-\mathcal{L}_0^{kl})^{-1} \mathcal{J}_k \rho_0]}{\text{Tr}[\mathcal{J}_k \rho_0]},
\end{eqnarray}
where  $\mathcal{L}_0^{kl}$ and $\mathcal{L}_0^{lm}$ are remainders of the Liouvillian defined in the analogous way as in Eq.~\eqref{wkldis}. If subsequent waiting times are uncorrelated the joint WTD can be factorized in the following way: $w_{klm}(\tau_{kl},\tau_{lm})=w_{kl}(\tau_{kl}) w_{lm}(\tau_{lm})$ and $w_{klm}(s,z)=w_{kl}(s)w_{lm}(z)$, otherwise it is not possible. Such a factorization is often referred to as the renewal property~\cite{dasenbrook2015}. Using the joint WTD one can calculate the cross-correlation (covariance) of two successive waiting times
\begin{eqnarray} \label{crosscor}
\langle \Delta \tau_{kl} \Delta \tau_{lm} \rangle = \left \{ \frac{\partial }{\partial s} \frac{\partial }{\partial z} \log[w_{klm} (s,z)]\right\}_{s=0,z=0},
\end{eqnarray}
where $\Delta \tau_{ij}=\tau_{ij}- \langle \tau_{ij} \rangle$ is the deviation from the mean. It is also useful to consider the normalized cross-correlation
\begin{eqnarray} \label{ncc}
NCC=\frac{\langle \Delta \tau_{kl} \Delta \tau_{lm} \rangle}{\sqrt{\langle \Delta \tau_{kl}^2 \rangle \langle \Delta \tau_{lm}^2 \rangle}},
\end{eqnarray}
defined as the covariance divided by the product of standard deviations of waiting times. If the renewal assumption is fulfilled $NCC=0$.
	
If the jumps associated with the operator $\mathcal{J}_k$ are renewal processes, one can find a direct relation between WTD and the cumulant generating function of FCS~\cite{brandes2008, albert2011}, as well as between the cumulants of FCS and WTD~\cite{albert2011, budini2011}. For nonrenewal processes such simple relations do not hold~\cite{budini2011, albert2012}. In Markovian systems, like the ones considered in this paper, all elementary processes (i.e the transitions between single states in the Liouville space) are renewal ones and one can always find a relation between the full counting statistics and the waiting time distributions of such processes~\cite{brandes2008}. This contrasts with the systems studied in Refs.~\cite{dasenbrook2015, albert2012}, in which the nonrenewal behavior was a result of the non-Markovian system-reservoir dynamics. However, also in Markovian systems the nonrenewal statistics can arise when the jump operator is associated with at least two elementary transitions~\cite{cao2006, caycedo-soler2008, budini2010, osadko2011}. Such complex forms of the jump operator are often necessary to describe the experimentally observed processes, because in many situations the elementary transitions are not directly accessible due to the presence of hidden degrees of freedom.
	
Finally, I consider the second-order current correlation function $S(\tau)$ which, in contrast to the waiting time distribution, is defined for the continuous current measurements. I focus on the auto-correlation function -- the correlation of values of the current flowing through a single junction measured at two different moments of time. It is expressed in the following way~\cite{brandes2008, emary2012}:
\begin{equation} \label{2ndordcur}
S(\tau)=\left \langle \delta I(\tau+t) \delta I(t) \right \rangle_t,
\end{equation}
where $\delta I(t)$ is the deviation of the current from its average in the stationary state. Its Fourier transform is the finite-frequency noise power~\cite{brandes2008, emary2012}
	\begin{equation} \label{finfreqnoi}
	S(\omega)=\int_{-\infty}^{\infty} d\tau e^{i\omega \tau} S(\tau).
	\end{equation}
	For the unidirectional transport $S(\tau)$ can be expressed as~\cite{emary2012}
\begin{equation} \label{relg2s}
S(\tau)=\langle I \rangle^2 [g^{(2)}(\tau)-1],
\end{equation}
where $\langle I \rangle$ is a mean current and $g^{(2)}(\tau)$ is the second-order correlation function defined as~\cite{carmichael1989, emary2012}	
\begin{equation} \label{2ndordpardef}
g^{(2)}(\tau)=\frac{\langle P(t,t+\tau) \rangle_t}{\langle P(t) \rangle_t^2},
\end{equation}
where $P(t,t+\tau)$ is the joint probability density of two electron tunnelings (not necessarily subsequent) occurring at times $t$ and $t+\tau$ respectively, and $P(t)$ is a probability density of the single electron tunneling at the time $t$. For the stationary state $\langle P(t) \rangle_t =\langle I \rangle$. The Laplace transform of the function $g^{(2)}(\tau)$ is given by the expression~\cite{emary2012}
\begin{equation} \label{2ndordpar}
g^{(2)}(s)=\frac{\text{Tr}[\mathcal{J} (s-\mathcal{L})^{-1} \mathcal{J} \rho_0]}{\text{Tr}[\mathcal{J} \rho_0]^2},
\end{equation}
where the operator $\mathcal{J}$ corresponds to all tunneling processes contributing to the measured current. In contrast to Eq.~\eqref{wkldis}, the above expression contains full Liouvillian $\mathcal{L}$ instead of the remainder $\mathcal{L}_0$.

For the Markovian renewal dynamics, $g^{(2)}(\tau)$ and $S(\tau)$ are related to the waiting time distribution $w(\tau)$. After some algebra, presented in Ref.~\cite{carmichael1989}, one obtains
\begin{eqnarray} \label{wg2s}
&& w(s) =\frac{\langle I \rangle g^{(2)}(s)}{1+\langle I \rangle g^{(2)}(s)}=\frac{\langle I \rangle^2 +s S(s)}{\langle I \rangle^2 +s [\langle I \rangle+S(s)]}.
\end{eqnarray}
Using this equality one can reconstruct the WTD on the basis of the second-order current correlation function. If the dynamics is nonrenewal, this relation is no longer valid~\cite{emary2012}. However, one can still use Eq.~\eqref{wg2s} to define the pseudo-waiting time distribution $u(\tau)$ and calculate its cumulants $q_n$ in the following way:
\begin{equation} \label{kumomega}
q_n= (-1)^n \left \{ \frac{d^n \log[u(s)]}{ds^n} \right\}_{s=0}.
\end{equation}
As shown below, this fictitious distribution may be useful for revealing the nonrenewal dynamics of the system.
	
%%%%%%%%%%%%%%%%%%%%%%%%%%%%%%%%%%%%%%%%%%%%%%%%%%%%%%%%%%%%%%%%%%%%%%%%%%%%%%%%%%%%%%%%%%%%%%%%%%%%%%%%%%%%%%%%%%%%%%%%%%%%%%%%%%%
%%%%%%%%%%%%%%%%%%%%%%%%%%%%%%%%%%%%%%%%%%%%%%%%%%%%%%%%%%%%%%%%%%%%%%%%%%%%%%%%%%%%%%%%%%%%%%%%%%%%%%%%%%%%%%%%%%%%%%%%%%%%%%%%%%%
	
\section{\label{sec:twodots}Two capacitively coupled dots}
\subsection{\label{subsec:model}Model}
\begin{figure}
		\centering
		\subfloat[]{\includegraphics[width=0.57\linewidth]{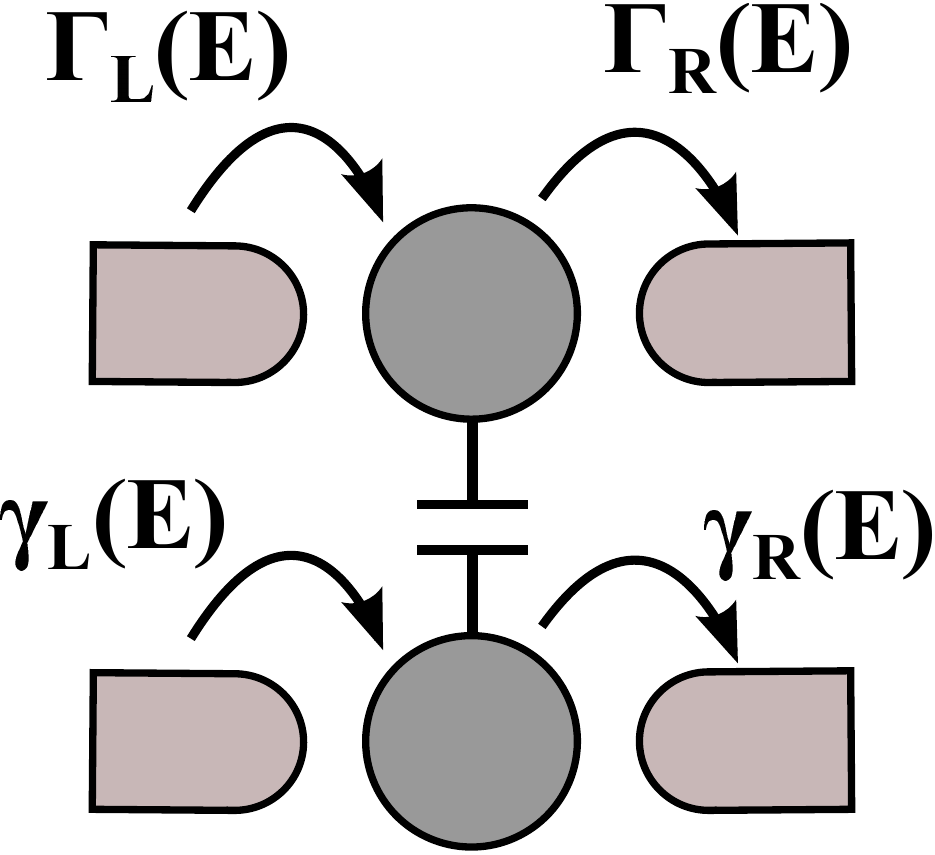}}
		\subfloat[]{\includegraphics[width=0.43\linewidth]{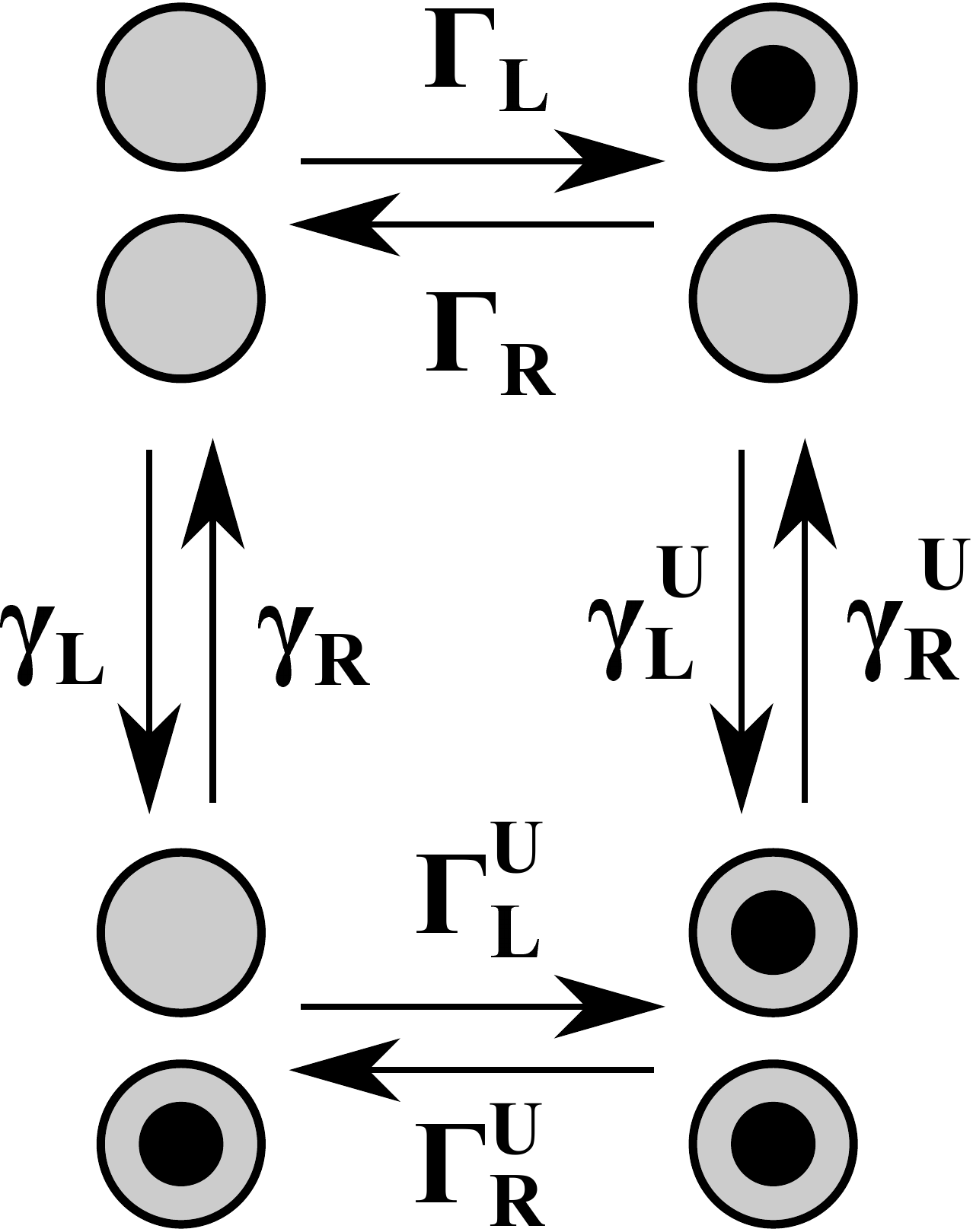}}
		\caption{(a) System of two capacitively coupled quantum dots; $\Gamma_i (E)$ and $\gamma_i(E)$ -- energy dependent tunneling rates. (b) Four-state Markovian model of the dynamics of the system. Each state corresponds to a different charge configuration, with black dots denoting the electrons occupying quantum dots. Transition rates with superscripts~$^U$ refer to the cases when the other dot is charged.}
		\label{fig:dqd}
	\end{figure}

In this section I consider the system of two capacitively coupled quantum dots [Fig.~\ref{fig:dqd}~(a)] to illustrate the generation of nonrenewal current fluctuations due to the telegraphic switching. Such a double quantum dot system, despite its simplicity, exhibits a rich physics, and therefore has been already extensively studied both experimentally~\cite{mcclure2007} and theoretically~\cite{michalek2009}. Previous studies have revealed, for example, phenomena like a Coulomb drag~\cite{sanchez2010, kaasbjerg2016, keller2016}, heat-to-current conversion~\cite{sanchez2011, thierschmann2015} and even Maxwell's-demon-like behavior~\cite{strasberg2013} (confirmed experimentally by using the system of two metallic islands~\cite{koski2015}). Moreover, recently the influence of the telegraphic switching on the full counting statistics in a similar system of two capacitively coupled single electron boxes has been experimentally investigated~\cite{singh2016}. 
	
The studied system is a four terminal one -- each dot is coupled to separate left (source) and right (drain) leads. As mentioned in Sec.~\ref{sec:methods}, the infinite-bias limit is taken. Tunneling rates between the dot and the leads are assumed to be energy-dependent, which is experimentally feasible~\cite{thierschmann2015}, and thus their values are conditioned on the occupancy of the other dot. Additionally, a strong intra-dot Coulomb repulsion is assumed -- each dot can be occupied by at most one electron. Therefore, the dynamics of the system can be described by a four-state Markovian model with states defined in the basis $\{ (0,0),(1,0),(0,1),(1,1)\}$, where the first and the second position in the brackets corresponds to the upper and the lower dot, while $0$ and $1$ refers to the empty and the occupied dot, respectively [Fig.~\ref{fig:dqd}~(b)]. The Liouvillian written in the considered basis reads
	\begin{eqnarray}
	\nonumber \mathcal{L}= 
	\begin{pmatrix}
	-\Gamma_L-\gamma_L & \Gamma_R & \gamma_R & 0 \\
	\Gamma_L & -\Gamma_R-\gamma_L^U & 0 & \gamma_R^U \\
	\gamma_L  & 0  & -\Gamma_L^U-\gamma_R & \Gamma_R^U  \\
	0 & \gamma_L^U & \Gamma_L^U & -\Gamma_R^U-\gamma_R^U
	\end{pmatrix}. \\
	\end{eqnarray}
	%%%%%%%%%%%%%%%%%%%%%%%%%%%%%%%%%%%%%%%%%%%%%%%%%%%%%%%%%%%%%%%%%
	%%%%%%%%%%%%%%%%%%%%%%%%%%%%%%%%%%%%%%%%%%%%%%%%%%%%%%%%%%%%%%%%
In principle, by using two separate detectors, it is possible to simultaneously detect charge states of both dots. In this way one can observe the whole dynamics of the system. Since the dynamics is Markovian, each transition between individual charge states is a renewal process. However, in this paper I focus on the statistics of tunneling events in the upper quantum dot only, treating the lower dot as an external source of a random telegraph noise (similarly to Refs.~\cite{schaller2010,singh2016}). The main reason for this is to provide a simple model which well illustrates the influence of telegraphic switching on current fluctuations (and, in particular, on waiting time correlations). Already in Refs.~\cite{cao2006, caycedo-soler2008, budini2010} it has been reported, that the telegraphic switching may generate a correlation between subsequent waiting times. Moreover, nonrenewal photon statistics in a similar four-state model have been already studied in Refs.~\cite{caycedo-soler2008, osadko2011}. However, this paper extends these previous studies in two ways. Firstly, it analyses additional statistical quantities, like the second and the third cumulant of current fluctuations. Secondly, it presents qualitatively new phenomena resulting from the reduction of symmetry of the model studied.
	
To explain, why the random switching is expected to occur in the system, the Liouvillian is decomposed as follows:
\begin{eqnarray}
\mathcal{L}=\mathcal{L}_{\Gamma}+\mathcal{L}_{\gamma},
\end{eqnarray}
where $\mathcal{L}_{\Gamma}$/$\mathcal{L}_{\gamma}$ are operators describing tunneling in the upper/lower dot (containing, respectively, tunneling rates $\{\Gamma_i, \Gamma_i^U \}$ and $\{\gamma_i, \gamma_i^U \}$). The operator $\mathcal{L}_{\Gamma}$ can be written in the block-diagonal form
\begin{eqnarray}
\mathcal{L}_{\Gamma}=\begin{pmatrix}
\mathcal{L}_{11} & 0 \\
0 & \mathcal{L}_{22} 
\end{pmatrix},
\end{eqnarray}
where matrices $\mathcal{L}_{11}$ and $\mathcal{L}_{22}$ contain transitions rates $\Gamma_i$ and $\Gamma_i^U$, respectively. These matrices describe tunneling in two distinct transport channels corresponding to different charge configurations of the lower dot. The operator $\mathcal{L}_{\gamma}$, meanwhile, describes the switching between these channels. Thus, the system is expected to exhibit the telegraphic switching behavior. 

In order to calculate the statistics of current fluctuations, I also define two jump operators
	\begin{eqnarray}
	\mathcal{J}_L 
	=\begin{pmatrix}
	0 & 0 & 0 & 0 \\
	\Gamma_L & 0 & 0 & 0 \\
	0  & 0  & 0 & 0  \\
	0 & 0 & \Gamma_L^U & 0
	\end{pmatrix}, \\ 
	\mathcal{J}_R 
	=\begin{pmatrix}
	0 & \Gamma_R & 0 & 0 \\
	0 & 0 & 0 & 0 \\
	0  & 0  & 0 & \Gamma_R^U  \\
	0 & 0 & 0 & 0
	\end{pmatrix},
	\end{eqnarray}
describing the tunneling from the left lead to the upper dot and the tunneling from the upper dot to the right lead, respectively. As one can note, each of these operators corresponds to two transitions within the four state model. As mentioned in Sec.~\ref{sec:methods}, the waiting time correlations can occur only, when the jump operator corresponds to at least two elementary transitions.
	
\subsection{\label{subsec:results2dots}Results}
	
%%%%%%%%%%%%%%%%%%%%%%%%%%%%%%%%%%%%%%%%%%%%%%%%%%%%%%%%%%%%%%%%%%%%
\begin{figure}
		\centering
		\subfloat[]{\includegraphics[width=0.348\linewidth]{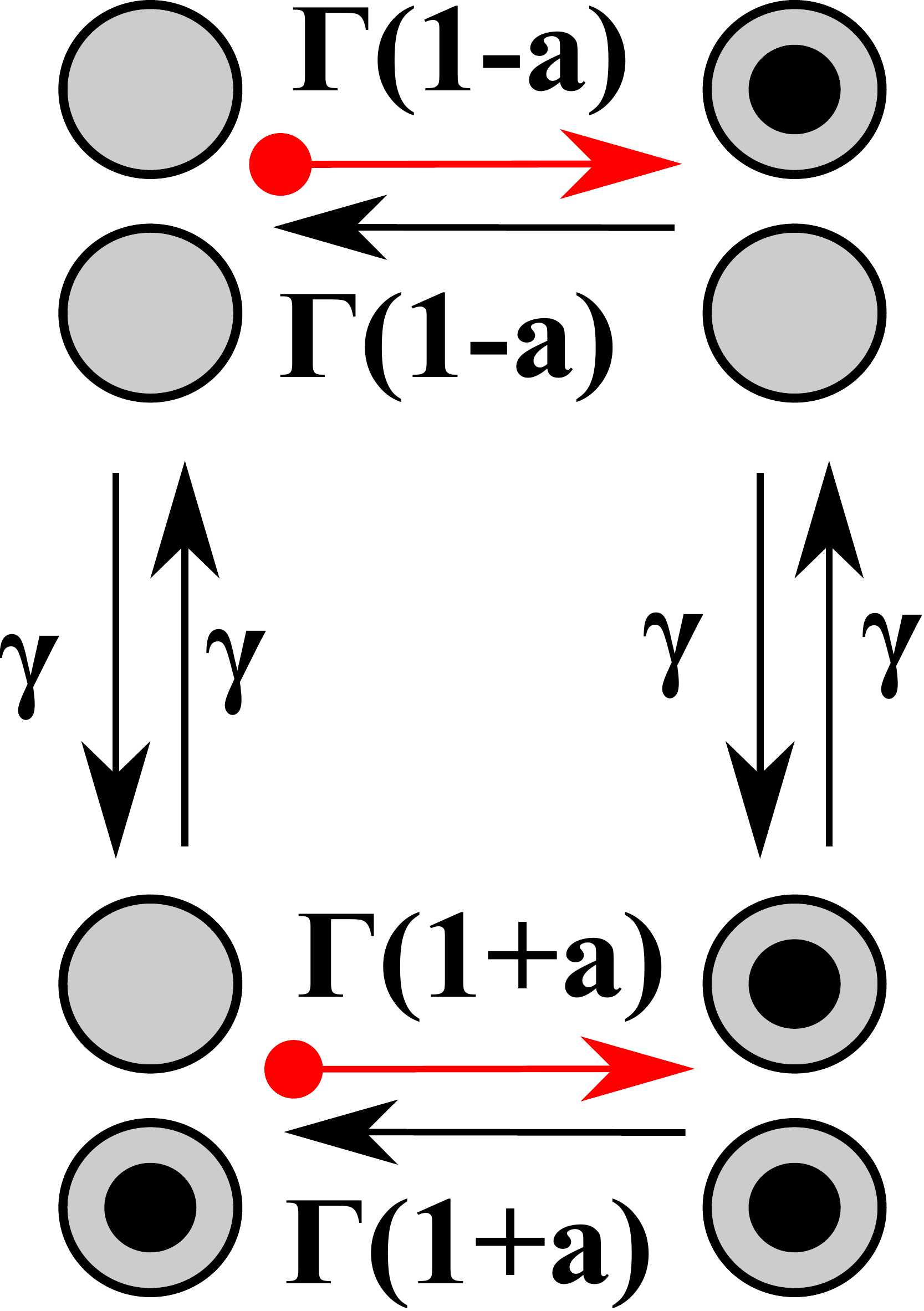}}
		\hspace{8mm}
		\subfloat[]{\includegraphics[width=0.398\linewidth]{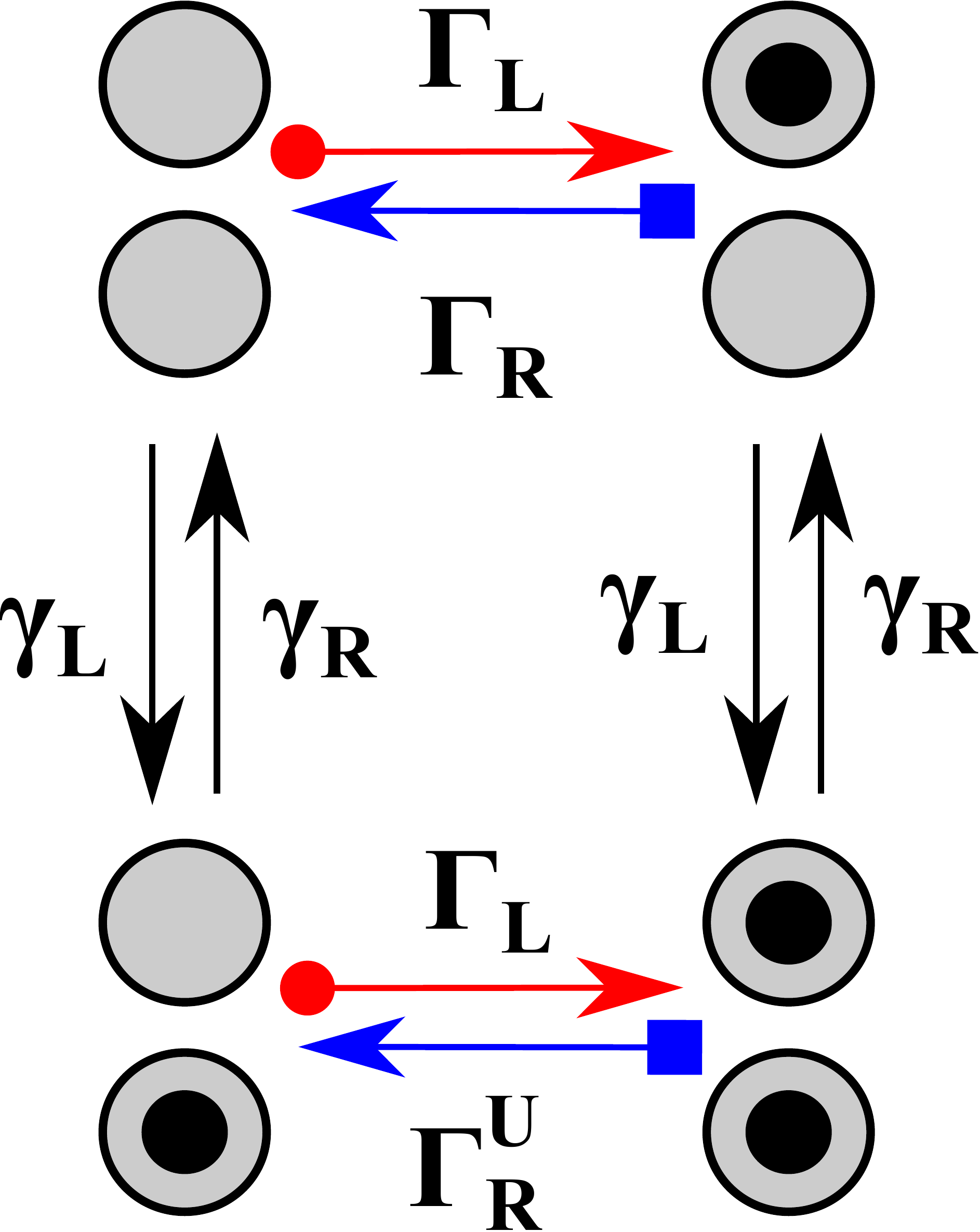}}
		\caption{(Color online). Four state models for different tunneling rate symmetries considered in the main text: (a) the symmetric system with $\gamma_i=\gamma_i^U=\gamma$, $\Gamma_i=(1-a) \Gamma$ and $\Gamma_i^U=(1+a) \Gamma$; (b) the asymmetric system with $\gamma_i=\gamma_i^U$ and $\Gamma_L^U=\Gamma_L$. Arrows with shaped tails correspond to counted transitions, the red arrows with bullet-tails correspond to the tunneling from the left lead to the dot, and blue arrows with square-tails in (b) correspond to the tunneling from the dot to the right lead.}
		\label{fig:dqdstates}
\end{figure}
%%%%%%%%%%%%%%%%%%%%%%%%%%%%%%%%%%%%%%%%%%%%%%%%%%%%%%%%%%%%%%%%%%%%
	
I begin my analysis with the simplest case illustrating the telegraphic switching: the system with dots symmetrically coupled to left and rights leads [Fig.~\ref{fig:dqdstates}~(a)]. The tunneling rates in the lower channel are assumed to be energy independent and equal to $\gamma$. The tunneling rates between the upper dot and the leads are equal to $(1-a) \Gamma$ when the lower dot is empty and $(1+a) \Gamma$ when the lower dot is occupied, where $a$ is the asymmetry parameter describing the energy dependence of tunneling rates.
	
%%%%%%%%%%%%%%%%%%%%%%%%%%%%%%%%%%%%%%%%%%%%%%%%%%%%%%%%%%
\begin{figure}
		\centering
		\subfloat{\includegraphics[width=0.96\linewidth]{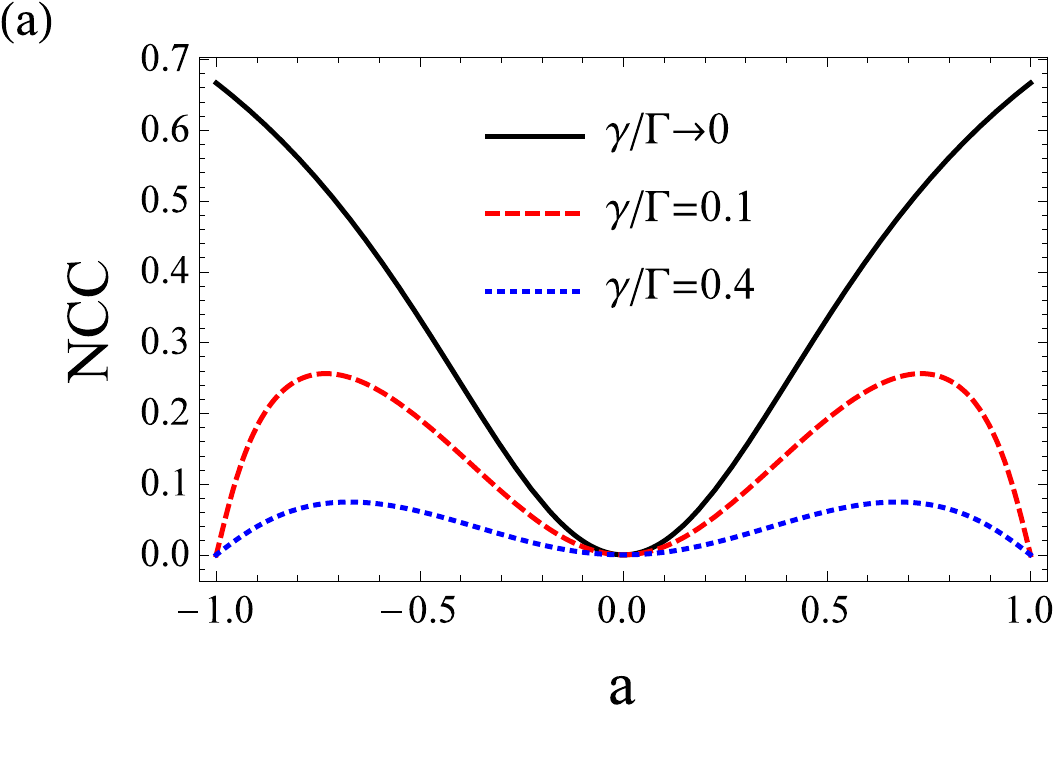}} \\
		\subfloat{\includegraphics[width=0.96\linewidth]{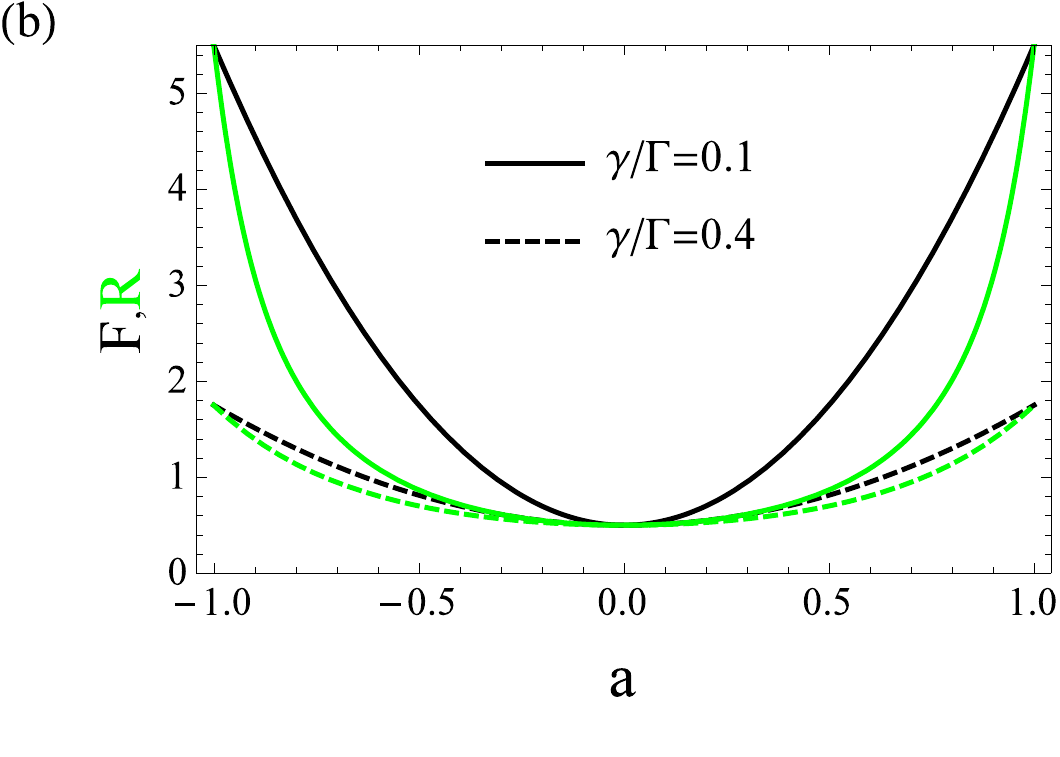}}
		\caption{(Color online). Dependence of (a) the normalized cross-correlation $NCC$ of successive waiting times $\tau_{LL}$, (b) the Fano factor $F$ (black curves) and the randomness parameter $R$ (green curves) on $a$ for the symmetric system [Fig.~\ref{fig:dqdstates}~(a)] for different values of $\gamma/\Gamma$.}
		\label{fig:resdqd1}
\end{figure}
%%%%%%%%%%%%%%%%%%%%%%%%%%%%%%%%%%%%%%%%%%%%%%%%%%%%%%%%%%
	
At first, I show that waiting time correlations occur in the system. I denote the waiting time between subsequent electron jumps from the left lead to the upper dot as $\tau_{LL}$. Fig.~\ref{fig:resdqd1}~(a) shows the dependence of the normalized cross-correlation of two subsequent times $\tau_{LL}$ on the asymmetry parameter $a$ for different ratios $\gamma/\Gamma$. One can observe that for $a \neq 0$ the cross-correlation can be positive, which clearly indicates that the dynamics is nonrenewal. As a matter of fact, the joint distribution of two subsequent waiting times cannot be factorized into a product of two single-waiting time distributions. The presence of the correlation results from the telegraphic switching between transport channels with different values of tunneling rates [either $(1+a) \Gamma$ or $(1-a) \Gamma$]. When the channel with higher tunneling rates (later referred to as the fast channel) is open, the subsequent waiting times tends to be shorter than the mean value. Accordingly, when the second (slow) channel is open, the subsequent waiting times tends to be longer than the mean value. As a result, the subsequent waiting times become positively correlated.

The cross-correlation is reduced as the ratio  $\gamma/\Gamma$ rises, because the fast switching between channels increases the probability that a short waiting time is followed by a long one and conversely. In the limiting case of $\gamma \to \infty$, the correlation vanishes. When the absolute value of $a$ is increased, the cross-correlation at first rises because the deviation of waiting times from the mean value increases in either channel. However, for high values of $|a|$ the NCC starts to decrease (expect for the limiting case of $\gamma \to 0$). In such a case the current flowing through a slow channel is blocked, and transport is predominated by the tunneling in the fast channel, which decreases the number of waiting times deviating from the mean value, and consequently also the value of cross-correlation. 
	
In the next step I analyze the influence of telegraphic switching on the full counting statistics and the waiting time distribution, and show that their joint analysis also provides information about the nonrenewal behavior of the system. It has been already reported by Albert~\textit{et al.}~\cite{albert2012}, who studied FCS and WTD in a quantum point contact. However, the system analyzed herein enables to study the qualitative behavior of FCS and WTD in a simpler way, and reveals also the specific physics of telegraphic switching phenomenon. I focus on the analysis of two quantities associated with the FCS and WTD respectively, the Fano factor $F$ and the randomness parameter $R$~\cite{chemla2008, moffitt2010}:
\begin{eqnarray}  
F&=&\frac{c_2}{c_1}=\lim_{t \to \infty} \frac{\langle \Delta n(t)^2 \rangle}{\langle n(t) \rangle}, \\
R&=&\frac{\kappa^{LL}_2}{(\kappa^{LL}_1)^2}=\frac{\langle \Delta \tau_{LL}^2 \rangle}{\langle \tau_{LL} \rangle^2},
\end{eqnarray}
where $n(t)$ is the number of particles tunneling through the upper dot within the time $t$. If the renewal assumption is satisfied, these parameters are the same~\cite{albert2011, budini2011}. As Fig.~\ref{fig:resdqd1}~(b) clearly shows, here this equality does not hold. In the considered system the Fano factor is always higher than or equal to the randomness parameter. The difference $F-R$ depends on $\gamma/\Gamma$ and $a$ in a similar way as the cross-correlation $NCC$. When the magnitude of the asymmetry parameter $|a|$ is sufficiently high, and at the same time the ratio $\gamma/\Gamma$ is low enough, the noise becomes super-Poissonian. Such a large noise enhancement resulting from the telegraphic switching is an already known phenomenon~\cite{koch2005, flindt2005, urban2009, sothmann2010, kaasbjerg2015}. In the limiting case of $\gamma \to 0$, the Fano factor becomes infinite for arbitrary non-zero values of $a$. The randomness parameter, however, remains finite also at this limit. Let's now explain this. For $\gamma \to 0$ the switching between transport channels becomes infinitesimally slow, and thus one can assume that during an arbitrarily long time interval either the fast or the slow channel is open. Therefore, one can consider two separate distributions of charge transmitted in the fast and the slow channel, denoted as $n_f(t)$ and $n_s(t)$, respectively. The variance of the number of transmitted particles is given by the following expression: 
\begin{eqnarray}
\nonumber \langle \Delta n(t)^2 \rangle &=& P_f \langle [n_f(t) - \langle n(t) \rangle]^2 \rangle+P_s \langle [n_s(t) - \langle n(t) \rangle]^2 \rangle 
\\ \nonumber &=&P_f \{\langle [n_f(t) - \langle n_f(t) \rangle]^2 \rangle+ [\langle n_f(t) \rangle - \langle n(t) \rangle]^2 \} 
\\ \nonumber &+&P_s \{\langle [n_s(t) - \langle n_s(t) \rangle]^2 \rangle+ [\langle n_s(t) \rangle - \langle n(t) \rangle]^2 \}, \\
\end{eqnarray}
where $P_f$ and $P_s$ denote the probabilities that the fast or the slow channel is open, respectively. For long times the variance is predominated by the sum of terms $[\langle n_j(t) \rangle - \langle n(t) \rangle]^2 =(I_j-I)^2 t^2$, where $j \in \{f,s\}$, $I$ is the mean current and $I_j$ is the mean current in the $j$-th channel. These terms are quadratic functions of time, while the mean value $\langle n(t) \rangle$ is a linear function of $t$, and thus $F=\lim_{t \to \infty} \langle \Delta n(t)^2 \rangle/\langle n(t) \rangle=\infty$. The randomness parameter, however, remains finite, which can be shown in a similar way. One can write the variance of waiting times as
\begin{eqnarray}
\nonumber \langle \Delta \tau^2 \rangle &=& p_f \langle (\tau_f - \langle \tau \rangle)^2 \rangle +p_s \langle (\tau_s - \langle \tau \rangle)^2 \rangle \\ 
&=& p_f [\langle (\tau_f - \langle \tau_f \rangle)^2 \rangle+  (\langle \tau_f \rangle-\langle \tau \rangle)^2 ] \\ \nonumber 
&+& p_s [\langle (\tau_s - \langle \tau_s \rangle)^2 \rangle+ (\langle \tau_s \rangle-\langle \tau \rangle)^2 ],
\end{eqnarray}
where $\tau_f/\tau_s$ are the waiting times measured in the fast/slow channel, while $p_f=P_f \langle \tau_s \rangle/(P_s \langle \tau_f \rangle +P_f \langle \tau_s \rangle)$ and $p_s=P_s \langle \tau_f \rangle/(P_s \langle \tau_f \rangle+P_f \langle \tau_s \rangle)$ are the probabilities that the charge is transmitted in the fast or the slow channel, respectively. It is a sum of finite quantities, and thus the randomness parameter remains finite.
	
As already mentioned in Sec.~\ref{sec:intro}, single electron counting experiments are currently confined to measurements of low currents corresponding to tunneling frequencies on the order of kHz. Moreover, they are not directly applicable for the study of electron transport beyond the sequential tunneling regime. However, it appears that the nonrenewal character of current fluctuations can be inferred using low-order current correlation functions, which in principle can be investigated by conventional current measurements. To show this, let's consider cumulants $q_n$ of the pseudo-waiting time distribution $u(\tau)$ calculated on the basis of the second-order current correlation function $S(\tau)$ (see the end of Sec.~\ref{sec:methods}). The pseudo-randomness parameter $\Pi=q_2/q_1^2$ appears to be equal to the Fano factor. However, it appears that the nonrenewal dynamics may be revealed by the behavior of higher order cumulants. Let's analyze the following quantities:
\begin{eqnarray} \label{skewnesses}
\sigma &=& \frac{c_3}{c_1}, \nonumber \\
\psi &=& 3 \frac{\kappa_2^2}{\kappa_1^4}-\frac{\kappa_3}{\kappa_1^3},\\
\xi &=& 3 \frac{q_2^2}{q_1^4}-\frac{q_3}{q_1^3}, \nonumber
\end{eqnarray}
%%%%%%%%%%%%%%%%%%%%%%%%%%%%%%%%%%%%%%%%%%%%%%%%%%%%%%%%%
	\begin{figure} 
		\centering
		\includegraphics[width=0.9\linewidth]{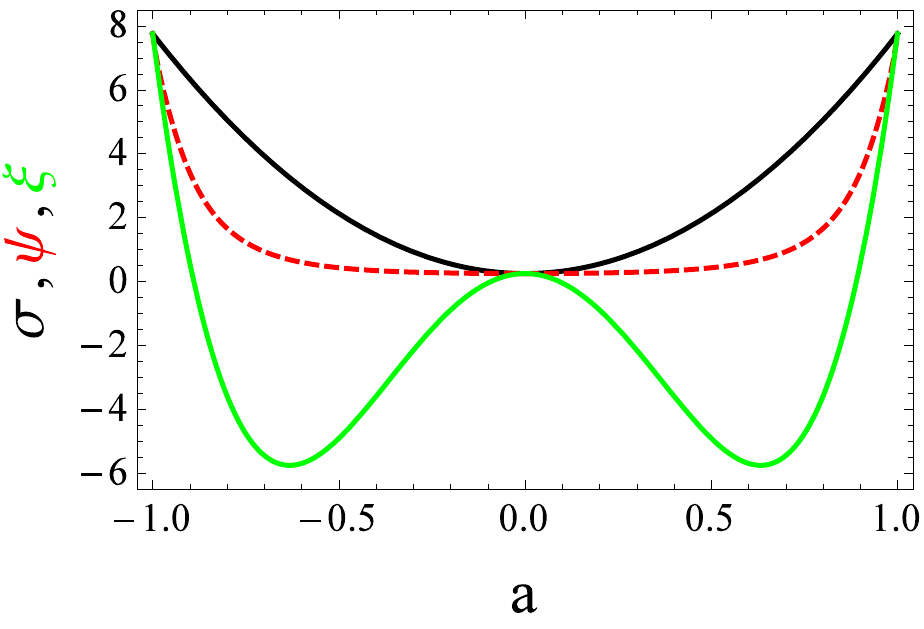} 
		\caption{(Color online). Dependence of the quantities defined in Eq.~\eqref{skewnesses} on $a$ for the symmetric system [Fig.~\ref{fig:dqdstates}~(a)] for $\gamma/\Gamma=0.1$. $\sigma$ -- black line, $\psi$ -- red dashed line, $\xi$ -- green line.}
		\label{fig:sksym}
	\end{figure}
%%%%%%%%%%%%%%%%%%%%%%%%%%%%%%%%%%%%%%%%%%%%%%%%%%%%%%%%%
where for simplicity $\kappa^{LL}_n=\kappa_n$. These three quantities are equal to each other for renewal systems~\cite{albert2011}, but in general they are unequal (see Fig.~\ref{fig:sksym}). Thus, it may be concluded that the joint measurement of the third cumulant of the zero-frequency FCS, which is experimentally accessible using continuous current measurements~\cite{bomze2005}, and the second-order current correlation function $S(\tau)$ enables to infer the presence of waiting time correlations. There is also a negative consequence: the identity~\eqref{wg2s}, which enables the reconstruction of WTD on the basis of $S(\tau)$ or $g^{(2)}(\tau)$, does not hold any longer. One should be aware that measurements of $S(\tau)$, though in principle possible~\cite{song2015, kobayashi2016}, are in general very demanding~\cite{ubbelohde2012}. However, I expect that experiments can give meaningful results at least for very slowly modulated systems. As a matter of fact, in such a case the step-like shape of the current-time trace could be even directly observed, as in Refs.~\cite{singh2016, lau2016}.
	
For the symmetric system the qualitative interpretation of NCC, WTD and FCS was relatively intuitive. Now I consider an asymmetric system [see Fig.~\ref{fig:dqdstates}~(b)], with different couplings of the dots to the left and right leads, to show that the joint dynamics of the upper and the lower dot may result in complex and non-trivial behavior of current fluctuations. As can be noted, now only one of tunneling rates is assumed to be energy dependent, i.e. the one between the upper quantum dot and the right lead (cf. a similar case in Ref.~\cite{sothmann2010}, in which a single tunneling rate was modulated by a spin impurity). 
	
\begin{figure}
		\centering
		\subfloat{\includegraphics[width=0.9455\linewidth]{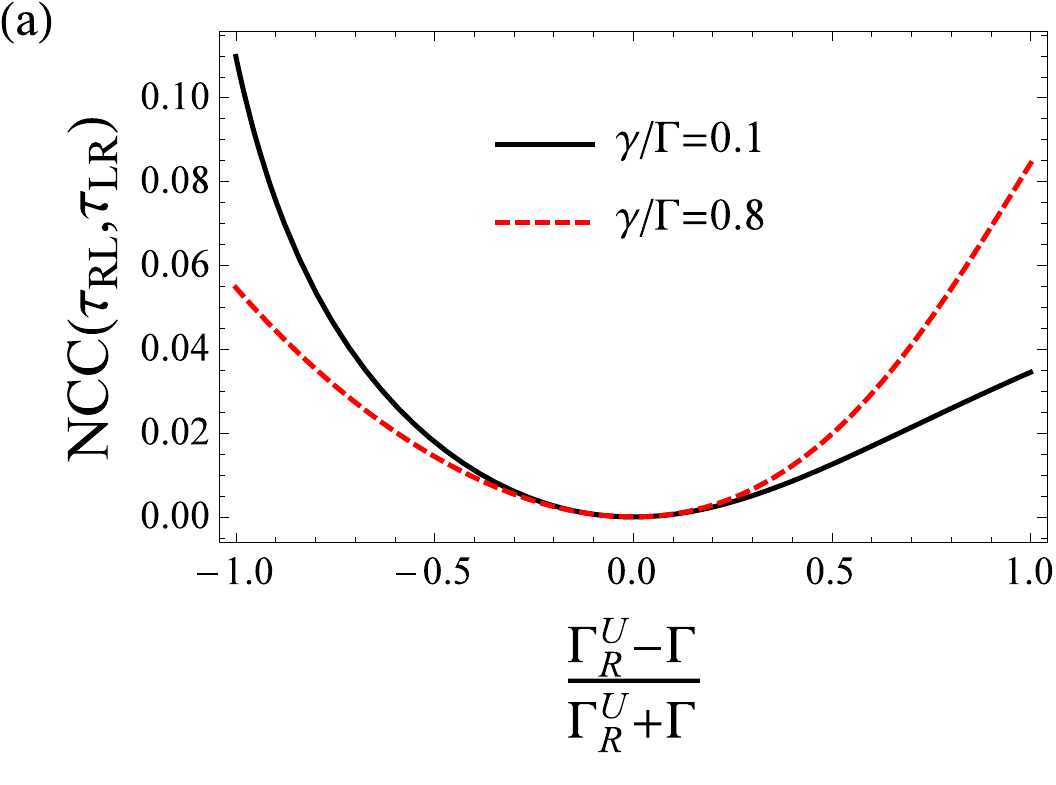}} \\
		\subfloat{\includegraphics[width=0.9455\linewidth]{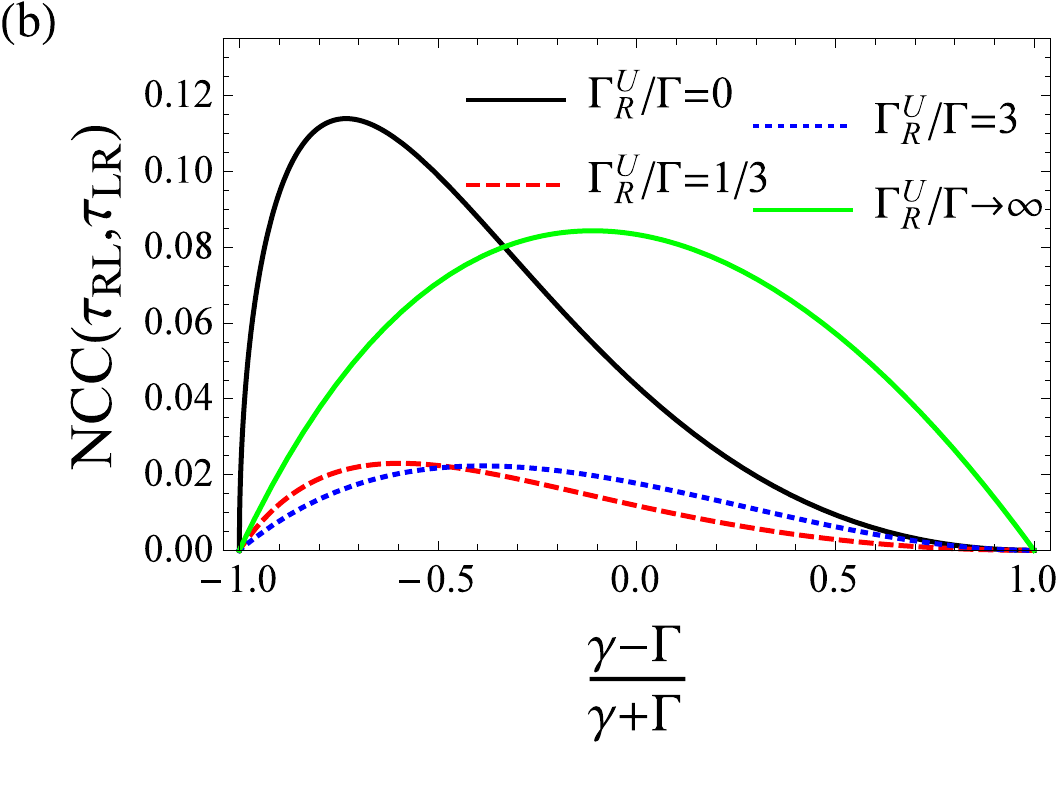}}
		\caption{(Color online). Normalized cross-correlation of successive times $\tau_{RL}$ and $\tau_{LR}$ for the asymmetric system [Fig.~\ref{fig:dqdstates}~(b)] with $\gamma_L=\gamma_R=\gamma$ and $\Gamma_L=\Gamma_R=\Gamma$: (a) dependence on $(\Gamma_R^U-\Gamma)/(\Gamma_R^U+\Gamma)$ for different values of $\gamma/\Gamma$; (b) dependence on $(\gamma-\Gamma)/(\gamma+\Gamma)$ for different values of $\Gamma_R^U/\Gamma$.}
		\label{fig:nccasym}
	\end{figure}

I begin the analysis by considering the correlation between subsequent times $\tau_{RL}$ (the time during which the dot remains empty) and $\tau_{LR}$ (the time during which the dot remains occupied -- the electron dwell time). Similar correlations in a double dot system, in which both tunneling rates in the upper dot were assumed to be modulated, have been studied previously in Ref.~\cite{ptaszynski2016}. As Fig.~\ref{fig:nccasym} shows, the modulation of a single tunneling rate is sufficient for generation of such correlations. This result may seem to be non-intuitive, since the rate of tunneling through the left lead, associated with the time $\tau_{RL}$, is now independent of the charge state of the lower dot. 

To unveil the mechanism generating the correlations, let's consider the qualitative behavior of NCC as a function of the system parameters. Firstly, as expected, the magnitude of the cross-correlation rises as the amplitude of the tunneling rate modulation $|\Gamma_R-\Gamma_R^U|$ increases [see Fig.~\ref{fig:nccasym}~(a)]. Let's now focus on the influence of the dynamics of the lower dot. For the sake of simplicity, I assume that $\gamma_L=\gamma_R=\gamma$. As follows, in contrast to the symmetric system, the cross-correlation depends on $\gamma$ in a non-monotonic way and reaches its maximum for some finite value of $\gamma$ [see Fig.~\ref{fig:nccasym}~(b)]. Moreover, for $\gamma \rightarrow 0$ the cross-correlation vanishes, which suggests that the correlation of waiting times is a result of correlated tunneling processes in the upper and the lower dot. Let's consider this in detail and analyze the following sequence of events: electron jump from the dot to the right lead -- waiting time $\tau_{RL}$ -- jump from the left lead to the dot -- waiting time $\tau_{LR}$ -- the second jump from the dot to the right lead. Two cases are considered: (i) when the time $\tau_{RL}$ is short and (ii) when it is long. For simplicity, I assume that $\Gamma_R^U>\Gamma_R$. At first let's note that due to $\Gamma_R^U>\Gamma_R$, the state $(0,1)$ rather than $(0,0)$ would be generated at the beginning of the process. Thus, for the short time $\tau_{RL}$ [the case (i)] the probability that successive evolution of the systems follows the trajectory $\theta_1 =(0,1) \rightarrow (1,1) \rightarrow (0,1)$ is increased in comparison with the probability of the trajectory $\theta_2=(0,0) \rightarrow (1,0) \rightarrow (0,0)$. As a result, the time $\tau_{LR}$ tends to be shorter than the mean value since the trajectory $\theta_1$ is associated with the occupancy of the lower dot which, due to $\Gamma_R^U>\Gamma_R$, allows faster tunneling from the upper dot. In consequence, the short time $\tau_{RL}$ is correlated to the short successive time $\tau_{LR}$. On the other hand, for long times $\tau_{RL}$ [the case (ii)] the initial imbalance of probabilities of states $(0,1)$ and $(0,0)$ is reduced due to the tunneling through the lower dot. Therefore, the probabilities of the state $(0,0)$ and the trajectory $\theta_2$ rise up, and the subsequent time $\tau_{LR}$ tends to be longer than the mean value. Considering both cases together, it is apparent that subsequent times $\tau_{RL}$ and $\tau_{LR}$ are positively correlated.
	
	%%%%%%%%%%%%%%%%%%%%%%%%%%%%%%%%%%%%%%%%%%%%%%%%%%%%%%%%%%%%%%%%%%%%%%%%%
	%%%%%%%%%%%%%%%%%%%%%%%%%%%%%%%%%%%%%%%%%%%%%%%%%%%%%%%%%%%%%%%%%%%%%%%%%
	\begin{figure} 
		\centering
		\hspace{4.3mm}
		\includegraphics[width=0.8845\linewidth]{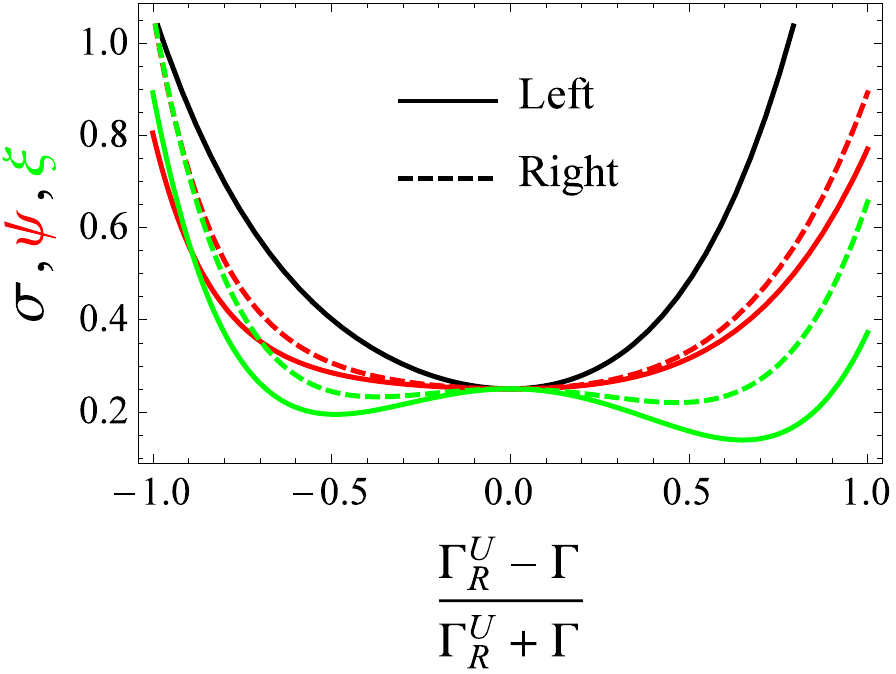} 
		\caption{(Color online). Dependence of the quantities defined in Eq.~\eqref{skewnesses} on $(\Gamma_R^U-\Gamma)/(\Gamma_R^U+\Gamma)$ for the asymmetric system [Fig.~\ref{fig:dqdstates}~(b)] with $\gamma_L=\gamma_R=\gamma$ and $\Gamma_L=\Gamma_R=\Gamma$ for $\gamma/\Gamma=0.3$. $\sigma$ -- black line, $\psi$ -- red line, $\xi$ -- green line. Continuous and dashed lines correspond to the quantities measured at the left and the right lead respectively.}
		\label{fig:skasym}
	\end{figure}
	%%%%%%%%%%%%%%%%%%%%%%%%%%%%%%%%%%%%%%%%%%%%%%%%%%%%%%%%%%%%%%%%%%%%%%%%%
	
One may ask, if the same waiting times ($\tau_{RL}$ and $\tau_{LR}$), but measured in the reversed order, are also correlated. It appears, that they are not. As can be noted, in the considered case the tunnel coupling of the upper dot to the left lead is the same for either the $(0,0)$ or the $(0,1)$ state. Thus, the rate of this process is in no way correlated to the previous course of the system's dynamics. As a result, subsequent times $\tau_{LR}$ and $\tau_{RL}$ are not correlated ($\langle \Delta \tau_{LR} \Delta \tau_{RL} \rangle=0$). Therefore, in general the correlation of two arbitrary waiting times $\tau_{ij}$ and $\tau_{ji}$ depends on their order: $\langle \Delta \tau_{ij} \Delta \tau_{ji} \rangle \not \equiv \langle \Delta \tau_{ji} \Delta \tau_{ij} \rangle$.
	
It has a direct consequence, which may also be surprising: the waiting time distributions measured at the left and the right junction, $w_{LL}(\tau)$ and $w_{RR}(\tau)$, can be in general unequal. To illustrate this, let us express the variances of waiting times measured at the left and the right junction ($\tau_{LL}$ and $\tau_{RR}$ respectively) as follows: $\langle \Delta \tau_{LL}^2 \rangle = \langle \Delta \tau_{LR}^2 \rangle+\langle \Delta \tau_{RL}^2 \rangle+2\langle \Delta \tau_{LR} \Delta \tau_{RL} \rangle $ and $\langle \Delta \tau_{RR}^2 \rangle = \langle \Delta \tau_{LR}^2 \rangle+\langle \Delta \tau_{RL}^2 \rangle+2\langle \Delta \tau_{RL}  \Delta \tau_{LR} \rangle$. As one can easily observe, these expressions differ only in the cross-correlation terms, and since these terms can be unequal, the variances can differ as well.
	
It also appears, that the functions $g^{(2)}(\tau)$, $S(\tau)$ and $S(\omega)$ measured at different junctions are also nonequivalent. At the same time, the mean currents and zero-frequency FCSs at both junctions are always the same, which is a consequence of the Kirchoff's current law. The pseudo-randomness parameter $\Pi=q_2/q_1^2$, measured at any junction, is equal to the Fano factor. However, the quantities $\psi$ and $\xi$ defined in Eq.~\eqref{skewnesses} are junction-dependent, as shown in Fig.~\ref{fig:skasym}. It should be mentioned, that a similar difference in the time-dependent factorial cumulants for incoming and outgoing electrons has been recently reported by Stegmann and K\"{o}nig~\cite{stegmann2016}.
	
Statistics of current fluctuations can be applied for the reconstruction of generators of dynamics of the systems with hidden degrees of freedom. For example, Bruderer~\textit{et al.}~\cite{bruderer2014} have proposed a procedure which enables determination of $n$ unknown parameters of the Liouvillian using zero-frequency FCS cumulants up to the $n$-th order. Although measurement of FCS cumulants up to the 15th order by single-electron counting techniques have been reported~\cite{flindt2009}, such experiments are in general very demanding~\cite{bruderer2014}. However, for nonrenewal systems the joint use of FCS and WTD may reduce the order of cumulants required for the reconstruction of the Liouvillian thanks to the independence of FCS and waiting time distributions measured at different junctions. For example, all parameters of the Liouvillian of the analyzed asymmetric system can be determined using the set of five independent quantities: $\{ \langle \tau_{LR} \rangle, \langle \tau_{RL} \rangle, \langle \Delta \tau_{LR}^2 \rangle, \langle \Delta \tau_{RR}^2 \rangle, F \}$. All these quantities are related only to the first and the second order cumulants, while the reconstruction on the basis of FCS would require the use of cumulants up to the fifth order.
	
\subsection{\label{subsec:dynchan}Comparison with the dynamical channel blockade}
%
%%%%%%%%%%%%%%%%%%%%%%%%%%%%%%%%%%%%%%%%%%%%%%%%%%%%%%%%%%%%%%%%%%%%%%%%%%
\begin{figure}
		\centering
		\subfloat[]{\includegraphics[width=0.54\linewidth]{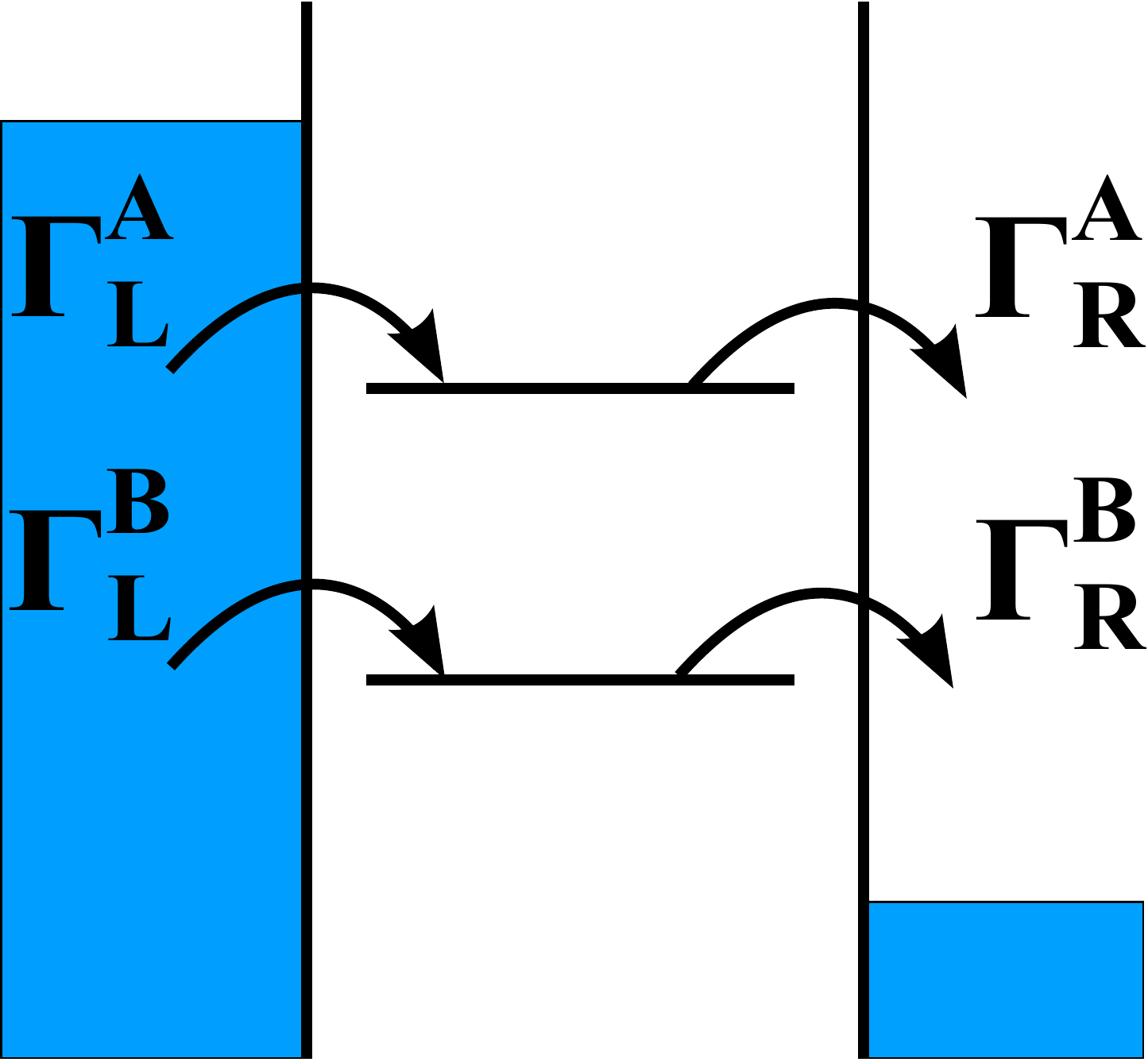}}
		\hspace{6mm}
		\subfloat[]{\includegraphics[width=0.16\linewidth]{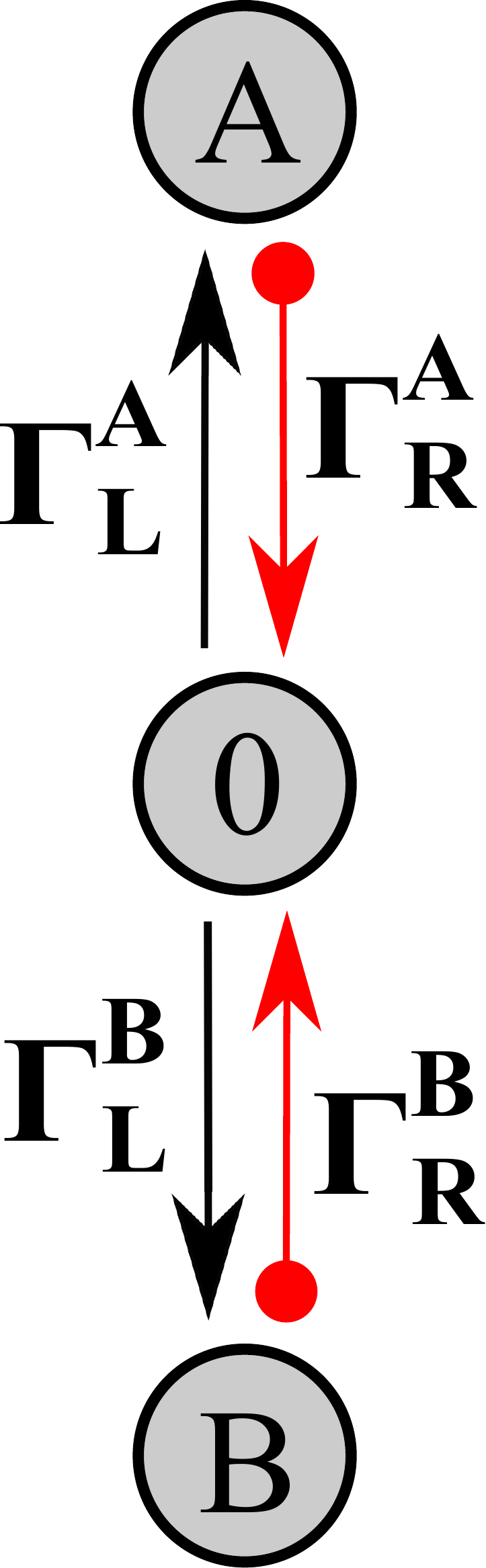}}
		\caption{(Color online). (a) Scheme of the dynamical channel blockade in a single quantum dot. (b) The three-state model of the dynamical channel blockade. Red arrows with bullet-tails correspond to the tunneling from the dot to the right lead.}
		\label{fig:dcb}
\end{figure}
%%%%%%%%%%%%%%%%%%%%%%%%%%%%%%%%%%%%%%%%%%%%%%%%%%%%%%%%%%%%%%%%%%%%%%%%%%
%
As implied by the results presented above, in the system exhibiting the telegraphic switching the noise enhancement can be associated with nonrenewal transport statistics (cf. Fig.~\ref{fig:resdqd1}; one should be aware that for $|a|=1$ the correlations vanish; thus, for example, the mechanism of noise enhancement reported by Safonov~\textit{et al.}~\cite{safonov2003} would not produce correlations). To highlight the significance of this fact, let's make a comparison with the other known mechanism of noise enhancement, namely the dynamical channel blockade, focusing on the most simple case -- a two-level quantum dot in the strong Coulomb blockade regime [Fig.~\ref{fig:dcb}~(a)]. The dot levels are assumed to be unequally coupled to the right lead. The infinite-bias limit is taken and the dynamics of the system is described by a three-state Markovian model, with tunneling through a single lead corresponding to two transitions between states [Fig.~\ref{fig:dcb}~(b)]. One can ask if this system exhibits the nonrenewal behavior. The answer is negative -- no correlation takes place. To explain this one should note an important difference between this model and the one considered previously. The double quantum dot examined beforehand is a multiple-reset system~\cite{brandes2008} -- there exist two sets of states corresponding to the specific charge configurations of the upper dot. Each tunneling event corresponds to the transition between these two sets. Because of the finite values of transition rates between the states within the single set, the departure trajectory from the set of states is correlated to the entry trajectory. The current system is a single-reset one~\cite{brandes2008} -- after each tunneling from the dot to the right lead the system returns to the same empty state and the successive trajectory is in no way correlated to the previous one. Thus, the dynamics of the system is renewal. This suggests that testing of the renewal property can be used as a tool to differentiate between different mechanisms leading to the noise enhancement, which can exhibit either the renewal or the nonrenewal dynamics.
	
%%%%%%%%%%%%%%%%%%%%%%%%%%%%%%%%%%%%%%%%%%%%%%%%%%%%%%%%%%%%%%%%%%%%%%%%%%%%%%%%%%%%%%%%%%%%%%%%%%%%%%%%%%%%%%%%%%%%%%%%%%%%%%%%%%%
%%%%%%%%%%%%%%%%%%%%%%%%%%%%%%%%%%%%%%%%%%%%%%%%%%%%%%%%%%%%%%%%%%%%%%%%%%%%%%%%%%%%%%%%%%%%%%%%%%%%%%%%%%%%%%%%%%%%%%%%%%%%%%%%%%%
	
\section{\label{sec:anderson}Anderson model}
So far, I have focused on the telegraphic switching phenomenon generating the positive cross-correlation of waiting times. However, this is by no means the only one mechanism producing nonrenewal current fluctuations. As a matter of fact, Caycedo-Soler~\textit{et al.}~\cite{caycedo-soler2008} have already investigated the three-level cascade-like optical system in which subsequent waiting times between photon emissions were shown to be negatively correlated. The case analyzed therein has a direct electronic analogue: the Anderson single impurity model in the infinite bias limit, which can be realized using a single quantum dot with a finite value of the intra-dot Coulomb interaction. Current fluctuations within this model have been previously investigated in Refs.~\cite{emary2012, brandes2008}, and Brandes~\cite{brandes2008} has even already mentioned its nonrenewal (multi-reset) character. In this section I go, however, beyond the previous studies by introducing spin polarization of leads attached to the dot, which generates the telegraphic switching. This enables us to observe the interplay of two mechanisms leading to either the negative or the positive cross-correlation of waiting times and analyze its influence on cumulants of FCS and WTD.
	
\subsection{\label{subsec:andmodel}Model}
	
The analyzed system [Fig.~\ref{fig:anderson}~(a)] consists of a quantum dot with a single electronic level. I consider unidirectional transport from the left to the right lead. The double occupancy of the dot by electrons with anti-parallel spins is now allowed. Thus, there are four states within the transport window: the empty state (denoted ${|0 \rangle}$), two singly-occupied states with different spin orientations (${|\! \uparrow\rangle}$ and ${|\! \downarrow\rangle}$) and the doubly occupied state (${|\! \uparrow \downarrow \rangle}$). Tunneling rates between the dot and the leads are assumed to be energy dependent, which makes them conditioned on the occupancy of the dot. As in the previously considered case, the energy-dependence of the tunneling rates is described by the asymmetry parameter $a$. For the sake of simplicity, I assume that the dot is symmetrically coupled to the left and the right lead. For spin-independent tunneling rates the dynamics of the system can be described by a three-state Markovian model [Fig.~\ref{fig:anderson}~(b)], with states $\{(0),(1),(2)\}$ referring to the empty, the singly-occupied and the doubly-occupied state, regardless of the spin. This is equivalent to the models investigated in Refs.~\cite{caycedo-soler2008, brandes2008, emary2012}. 
	
Moreover, the leads are assumed to be equally spin polarized with the polarization defined as $p={(\rho_{\uparrow}-\rho_{\downarrow})}/{(\rho_{\uparrow}+\rho_{\downarrow})}$, where $\rho_{\uparrow}$, $\rho_{\downarrow}$ are the densities of states for $\uparrow$ and $\downarrow$ electrons. The tunneling rates for different spin polarizations are then expressed as follows: $\Gamma_{\uparrow \pm}=(1+p)\Gamma_{\pm}$ and $\Gamma_{\downarrow \pm}=(1-p)\Gamma_{\pm}$, where $\Gamma_\pm=(1 \pm a) \Gamma$. The dynamics of the system is then described by a four-state Markovian model [Fig.~\ref{fig:anderson}~(c)]. The Liouvillian written in the basis $\{ {| 0 \rangle},{|\! \uparrow\rangle},{|\! \downarrow\rangle},{|\! \uparrow \downarrow \rangle} \}$ is expressed as follows:
	%%%%%%%%%%%%%%%%%%%%%%%%%%%%%%%%%%%%%%%%%%%%%%%%%%%%%%%%%%%%%%%%
	\begin{figure}
		\centering
		\subfloat[]{\includegraphics[width=0.55\linewidth]{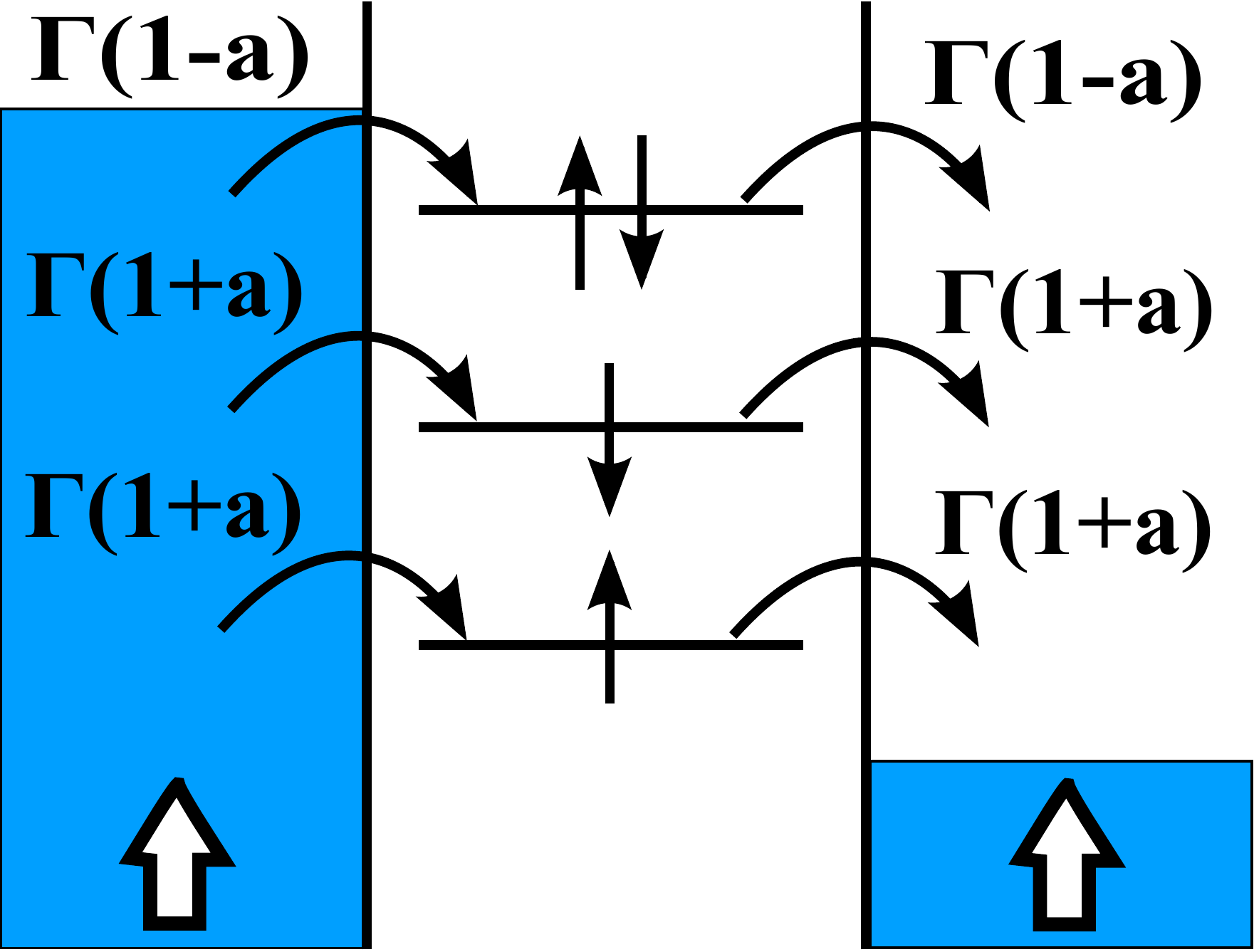}}
		\hspace{1mm}
		\subfloat[]{\includegraphics[width=0.32\linewidth]{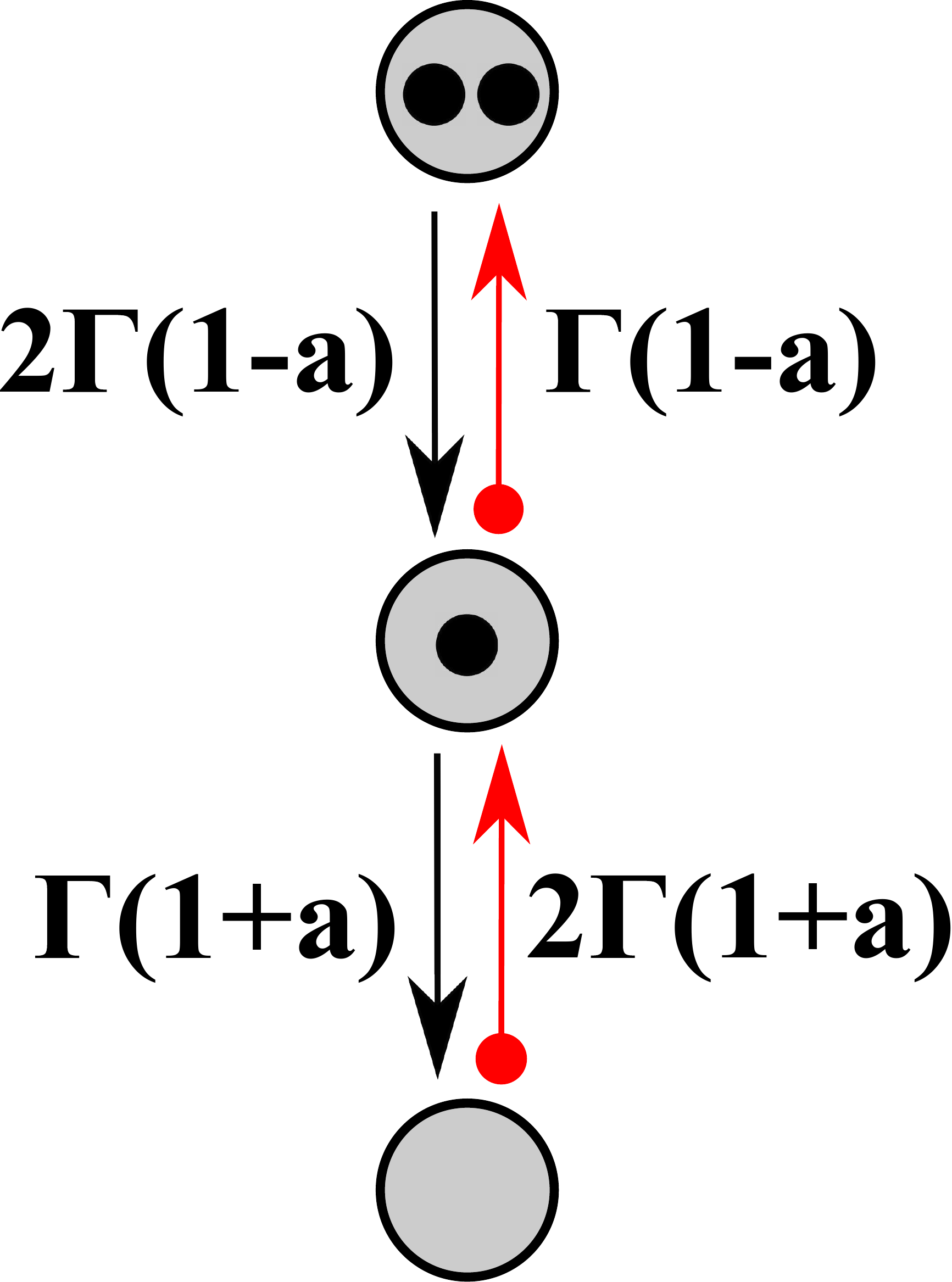}} \\
		\subfloat[]{\includegraphics[width=0.8\linewidth]{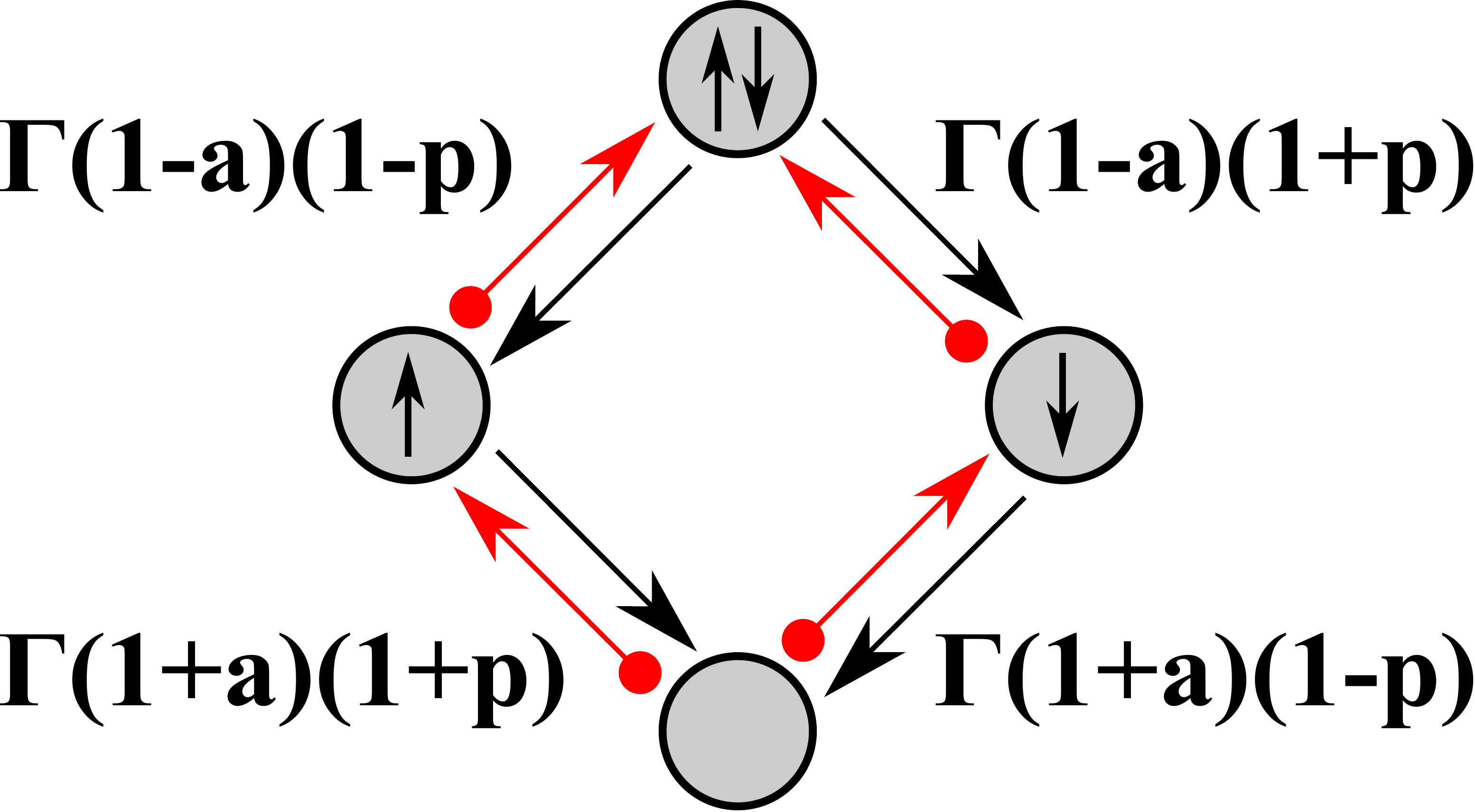}}
		\caption{(Color online). (a) Scheme of the transport within the Anderson model in the infinite bias limit. (b,c) Three- and four-state Markovian models of systems without and with spin polarization of the leads. Red arrows with bullet-tails denote transitions corresponding to the tunneling from the left lead to the dot.}
		\label{fig:anderson}
	\end{figure}
	%%%%%%%%%%%%%%%%%%%%%%%%%%%%%%%%%%%%%%%%%%%%%%%%%%%%%%%%%%%%%%%%
	\begin{eqnarray}
	\nonumber \mathcal{L}=
	\Gamma  \left(
	\begin{array}{cccc}
	-2 (a+1) & A^{(+)}P^{(+)} & A^{(+)}P^{(-)} & 0 \\
	A^{(+)}P^{(+)} & -2 a p-2 & 0 & A^{(-)}P^{(-)} \\
	A^{(+)}P^{(-)} & 0 & 2 a p-2 & A^{(-)}P^{(+)} \\
	0 & A^{(-)}P^{(-)} & A^{(-)}P^{(+)} & -2 (1-a) \\
	\end{array}
	\right), \\
	\end{eqnarray}
where $A^{(\pm)}=1 \pm a$, $P^{(\pm)}=1 \pm p$. Afterwards, I focus on the tunneling from the left lead to the dot. The operator counting all transitions in the junction, irrespectively of the electron spin and the occupancy of the dot, is introduced as
	\begin{eqnarray}
	\mathcal{J}_L= 
	\Gamma  \left(
	\begin{array}{cccc}
	0 & 0 & 0 & 0 \\
	A^{(+)}P^{(+)} & 0 & 0 & 0 \\
	A^{(+)}P^{(-)} & 0 & 0 & 0 \\
	0 & A^{(-)}P^{(-)} & A^{(-)}P^{(+)} & 0 \\
	\end{array}
	\right).
	\end{eqnarray}
In comparison with the model considered in the previous section [Fig.~\ref{fig:dqd}~(b)], the tunneling through a single junction corresponds herein to four, instead of two, different transitions between states. As shown below, the way in which electron tunnelings are counted (and thus the form of the jump operator) has a significant impact on results.
	
In contrast to the optical systems (like the one considered by Caycedo-Soler~\textit{et al.}~\cite{caycedo-soler2008}), in which emissions of photons are detected, in the electronic systems one can experimentally detect charge states, and thus analyze not only jumps through a single lead, but also elementary transitions $(0) \leftrightarrow (1)$ and $(1) \leftrightarrow (2)$. These elementary processes are renewal ones even in the case when the leads are spin-polarized. In view of this, the analysis of nonrenewal statistics may seem superfluous. However, Prance~\textit{et al.}~\cite{prance2015} have shown that the detection of electron jumps rather than charge states may be preferred due to the lesser vulnerability to noise. Therefore, the theoretical description applied in this paper has experimental relevance.
	
The analyzed model assumes that only the single level is available for the transport, and at the same time the zero, single and double occupancy of the dot is allowed. In real multi-level quantum dots this requires the intra-dot Coulomb interaction energy $U$ to be smaller than the separation between electronic levels of the dot. One should be aware that in relatively large lateral quantum dots realized in the two-dimensional electronic gas, where most of single-electron counting experiments have been conducted, the opposite is true -- the charging energy is much greater than the level spacing~\cite{matveev1996, kouwenhoven1997}. However, the considered model can be realized, for example, in ultrasmall quantum dots, in which the level spacing is high due to strong quantum confinement~\cite{shin2010}, or in three-dimensional CdSe or InAs dots, where $U$ can be smaller than the energy separation between the $s$- and $p$- orbitals due to the screening of the Coulomb interaction~\cite{bakkers2002, shibata2012}. Also the other experimental requirements can be satisfied in aforementioned systems, since the coupling of the InAs quantum dot to spin polarized leads has been achieved~\cite{hamaya2007}, and the single-electron counting in individual CdSe dots using carbon nanotube detectors has been realized~\cite{zdrojek2009, zbydniewska2015}. Thus, the experimental verification of the results presented herein by state-of-the-art techniques may be feasible.
	
\subsection{\label{subsec:andresults}Results}
\begin{figure} 
		\centering
		\includegraphics[width=0.9\linewidth]{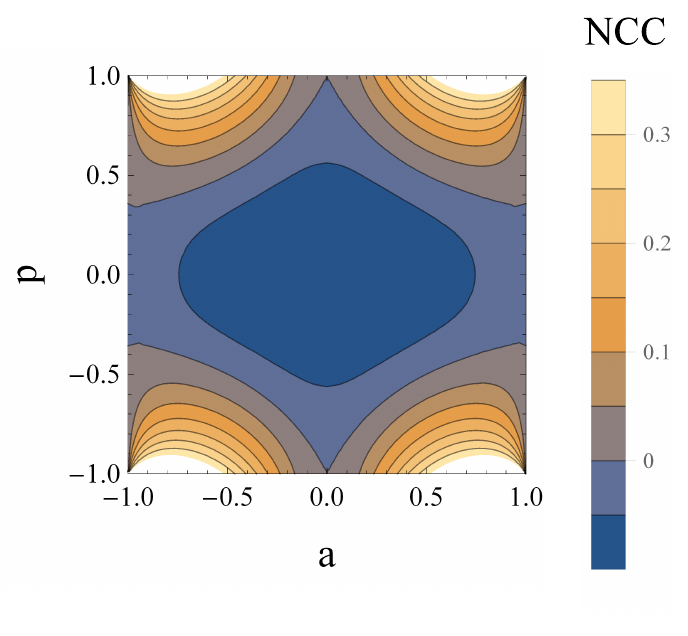} 
		\caption{(Color online). Dependence of the normalized cross-correlation $NCC$ of successive waiting times $\tau_{LL}$ on $a$ and $p$.}
		\label{fig:nccand}
	\end{figure}

\begin{figure}
		\centering
		\includegraphics[width=0.9\linewidth]{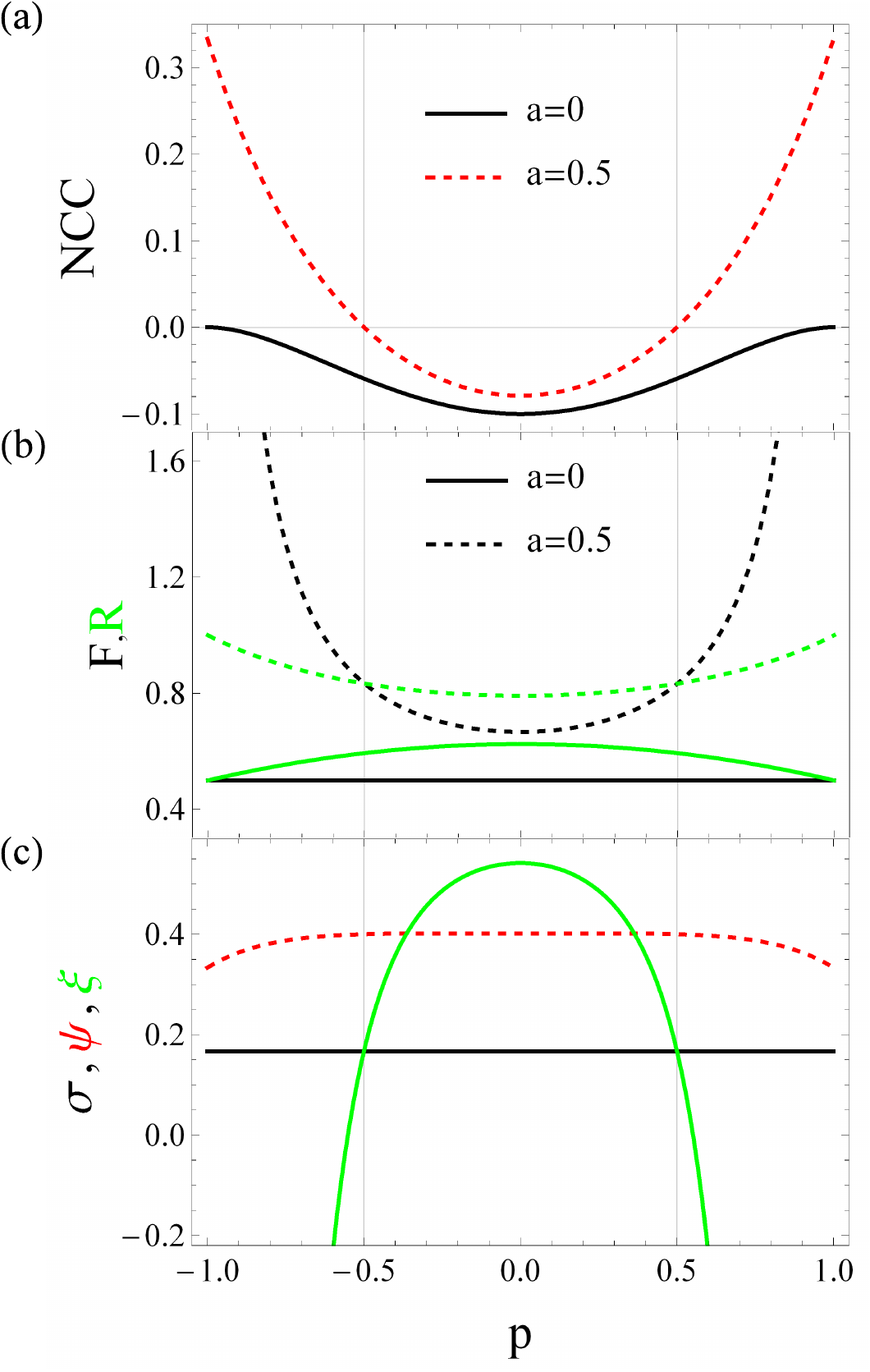} 
		\caption{(Color online). (a,b) Dependence of (a) the normalized cross-correlation $NCC$ of successive waiting times $\tau_{LL}$, (b) the Fano factor $F$ (black curves) and the randomness parameter $R$ (green curves) on $p$ for different values of $a$. (c) The plot of the quantities defined in Eq.~\eqref{skewnesses} vs $p$ for $a=0.5$. $\sigma$ -- black continuous line, $\psi$ -- red dashed line, $\xi$ -- green continuous line.}
		\label{fig:resand2}
	\end{figure}

I begin the analysis by considering the normalized cross-correlation of two successive times $\tau_{LL}$. The dependence of the cross-correlation on $a$ and $p$ is shown as a two dimensional contour plot in Fig.~\ref{fig:nccand}, with two cross-sections of this plot shown in Fig.~\ref{fig:resand2}~(a). It is apparent, that the cross-correlation can be both positive or negative, which contrasts with the case considered in Sec.~\ref{sec:twodots}, where only the positive cross-correlation occurred. This indicates that in the currently considered system there exist at least two competing processes leading to the negative and the positive cross-correlation, respectively. Let's first analyze the process leading to the negative cross-correlation, which is clearly different from the one considered previously. As the definition of the cross-correlation [Eq.~\eqref{crosscor}] indicates, the sign may become negative when the times positively and negatively deviating from the mean are intermingled. Le's now focus on the case of $p=0$. As follows from Fig.~\ref{fig:nccand}, the cross-correlation is then always negative. In this case the dynamics can be described by the three-state model [Fig.~\ref{fig:anderson}~(b)], later referred to as the cascade model. Let's now analyze the following trajectory of transitions between the states of the system: $(0) \rightarrow (1) \rightarrow (2)$. The waiting time between transitions $(0) \rightarrow (1)$ and $(1) \rightarrow (2)$ may be short -- these events may occur even simultaneously [$w(\tau=0)$ is finite]. However, if such a trajectory is realized, the next tunneling to the dot is blocked until at least one electron departs the dot, and therefore the waiting time tends to be longer than the mean one. In consequence, a short waiting time tends to be followed by a long one, which results in the negative cross-correlation. This can be illustrated by considering the difference $w(\tau_1,\tau_2)-w(\tau_1)w(\tau_2)$, where $w(\tau_1,\tau_2)$ is the joint distribution of two subsequent times $\tau_{LL}$ (denoted as $\tau_1$ and $\tau_2$), and $w(\tau_1)w(\tau_2)$ is the product of two single waiting times distributions. For the sake of better clarity of presentation I plot the normalized quantity
	%
	%%%%%%%%%%%%%%%%%%%%%%%%%%%%%%%%%%%%%%%%%%%%%%%%%%%%%%%%%%%%%%%%%%
	%%%%%%%%%%%%%%%%%%%%%%%%%%%%%%%%%%%%%%%%%%%%%%%%%%%%%%%%%%%%%%%%%%
	%%%%%%%%%%%%%%%%%%%%%%%%%%%%%%%%%%%%%%%%%%%%%%%%%%%%%%%%%%%%%%%%%%
	%%%%%%%%%%%%%%%%%%%%%%%%%%%%%%%%%%%%%%%%%%%%%%%%%%%%%%%%%%%%%%%%%%
	%
	%%%%%%%%%%%%%%%%%%%%%%%%%%%%%%%%%%%%%%%%%%%%%%%%%%%%%%%%%%%%%%%
	\begin{figure} 
		\centering
		\includegraphics[width=0.9\linewidth]{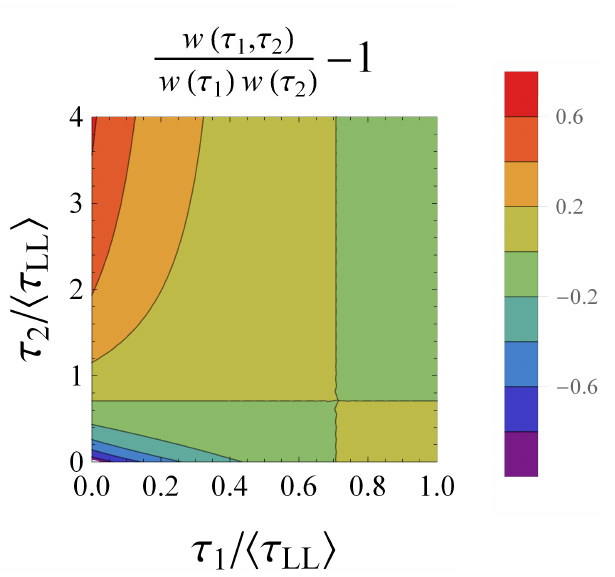} 
		\caption{(Color online). The normalized difference of the joint distribution of two successive waiting times $\tau_{LL}$ (denoted $\tau_1$ and $\tau_2$) and the product of two single-waiting time distributions for $a=0$ and $p=0$.}
		\label{fig:mapt1t2casc}
	\end{figure}
	%%%%%%%%%%%%%%%%%%%%%%%%%%%%%%%%%%%%%%%%%%%%%%%%%%%%%%%%%%%%%%%%
	%	
	\begin{eqnarray} \label{delta}
	\Delta(\tau_1,\tau_2)=\frac{w(\tau_1,\tau_2)}{w(\tau_1)w(\tau_2)}-1,
	\end{eqnarray}
which is the aforementioned difference divided by $w(\tau_1)w(\tau_2)$. The plot of $\Delta(\tau_1,\tau_2)$ for $p=0$ and $a=0$ is shown in Fig.~\ref{fig:mapt1t2casc}. As one can observe, for $\tau_1$ close to 0 the probability that the time $\tau_2$ is shorter than the mean time $\langle \tau_{LL} \rangle=1/\Gamma$ is decreased, and the probability that it is longer is increased, which confirms the qualitative reasoning presented above. Now, coming to the issue of the positive cross-correlation, let's consider the case when $a \neq 0$ and $p \approx 1$. At first, it should be noted that electrons with different spin polarizations flow through separate but mutually interacting transport channels. The main contribution to the current comes from the tunneling of the majority spins. The transport of the minority spins slowly modulates the conductance of the spin-majority channel due to the dependence of tunneling rates on the occupancy of the dot. Thus, one can observe the telegraphic switching process similar to the one occurring in the previously considered double dot system. As a result, the cross-correlation becomes positive. As Fig.~\ref{fig:nccand} indicates, in some cases the competition of the above considered processes makes the cross-correlation equal to 0.
	
In the next step I analyze the Fano factor and the randomness parameter. Their dependence on $p$ for two different values of $a$ is shown in Fig.~\ref{fig:resand2}~(b). For $a=0$ the telegraphic switching does not occur (the tunneling rates are independent of the occupancy of the dot) and the Fano factor is equal to 1/2 irrespectively of $p$. This finding may be explained as follows: there exist two independent transport channels for different spin polarizations. Each of them can be modeled as a symmetrically-coupled single-level dot in the strong Coulomb blockade regime, in which the Fano factor is equal to 1/2~\cite{bagrets2003}. In contrast, the randomness parameter is increased due to the cascade-like behavior, and reaches its maximum for $p=0$. For $a=0.5$ and $p \neq0$ the telegraphic switching occurs and both the Fano factor and the randomness parameter are increased. One can note a coincidence between the signs of the cross-correlation and the expression $F-R$. This coincidence was confirmed for different sets of the system parameters. In Sec.~\ref{subsec:results2dots} it was shown, why in the case of telegraphic switching the positive-cross correlation of waiting times is associated with the positive sign of the difference $F-R$. Let's now consider the case of negative cross-correlation using the following simplified model: the system emits electrons in a deterministic way, with the even/odd waiting times equal to $\tau_1$ or $\tau_2$ respectively. The cross-correlation of the subsequent waiting times is negative, because short and long waiting times are intermingled. Since the system is deterministic, in the long time limit the number of particles transferred in the time $t$ is not fluctuating and is given by the expression $n(t)=\langle n(t) \rangle=t/\langle \tau \rangle$, where $\langle \tau \rangle$ is the mean waiting time $(\tau_1+\tau_2)/2$. Thus, the Fano factor is equal to 0. The variance of the waiting times is however non-zero, because there are two different waiting times, and it is given by the following expression: $\langle \Delta \tau^2 \rangle = [(\tau_1-\langle \tau \rangle)^2+(\tau_2-\langle \tau \rangle)^2]/2=(\tau_1-\tau_2)^2/4$. Therefore, the randomness parameter is positive and $R>F$. Similar reasoning applied to the case, when waiting times are stochastically distributed, and at the same time two successive waiting times are in general negatively correlated, should explain the coincidence of signs of the cross-correlation and $F-R$ in the model considered. However, to check the generality of this rule I have also analyzed other Markovian models with different numbers of states and arbitrary chosen jump operators. The analysis has shown that the relation considered is not universal, but rather applies only to specific systems. It may be associated with the fact that the reasoning above neglects the correlations of higher numbers of successive waiting times [i.e. associated with distributions of the type $w(\tau_1,\tau_2,...,\tau_N)$].
	
%%%%%%%%%%%%%%%%%%%%%%%%%%%%%%%%%%%%%%%%%%%%%%%%%%%%%%%%%%%%%%%%
\begin{figure}
		\centering
		\subfloat{\includegraphics[width=0.88\linewidth]{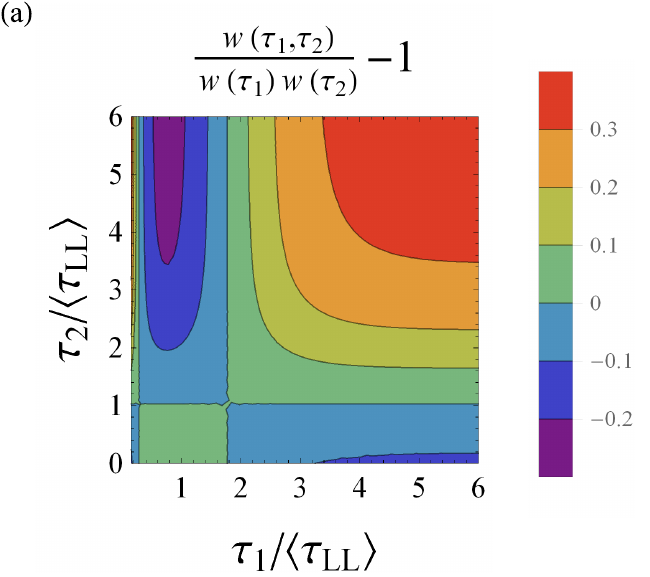}} \\
		\subfloat{\includegraphics[width=0.88\linewidth]{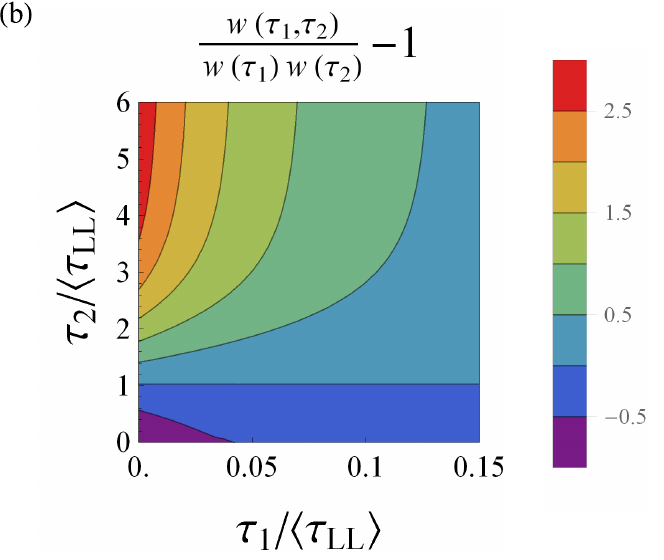}}
		\caption{(Color online). The normalized difference of the joint distribution of two successive waiting times $\tau_{LL}$ and the product of two single-waiting time distributions for $a=0.5$ and $p=0.5$ for different ranges of $\tau_1$: (a)~[0.15,~6.0] and (b)~[0.0,~0.15].}
		\label{fig:mapt1t2}
\end{figure}
%%%%%%%%%%%%%%%%%%%%%%%%%%%%%%%%%%%%%%%%%%%%%%%%%%%%%%%%%%%%%%%
	
Let's now consider the quantities related to the cumulants of the third order defined in Eq.~\eqref{skewnesses}. Their dependence on $p$ for a chosen value of the asymmetry parameter $a$ is shown in Fig.~\ref{fig:resand2}~(c). Again, one can note that the sign of the expression $\sigma-\xi$ is coincident with the sign of the cross-correlation [cf. Fig.~\ref{fig:resand2}~(a)]. Also this coincidence was confirmed for different sets of the parameters of the model considered. However, also in this case the analysis of other Markovian models has shown that this relation is not universal. The parameter $\psi$, related to the cumulants of WTD, is not equal to $\sigma$ even for $a=0.5$ and $p=0.5$, when the cross-correlation equals 0. This suggests that the dynamics is still nonrenewal -- the joint distributions of successive waiting times $w(\tau_1,\tau_2)$ cannot be factorized into two separate waiting time distributions. It can be shown by plotting the parameter $\Delta (\tau_1,\tau_2)$ defined in Eq.~\eqref{delta} (Fig.~\ref{fig:mapt1t2}). For sufficiently high values of $\tau_1$ the function $\Delta(\tau_1,\tau_2)$ is positive when both successive times are simultaneously short or long, while negative otherwise [see Fig.~\ref{fig:mapt1t2}~(a)], which is a result of the telegraphic switching. On the other hand, focusing on small values of $\tau_1$ [Fig.~\ref{fig:mapt1t2}~(b)] enables us to see a positive correlation of very short waiting times and the successive long ones, which is a trace of the cascade-like behavior (cf. Fig.~\ref{fig:mapt1t2casc}).
	
	%%%%%%%%%%%%%%%%%%%%%%%%%%%%%%%%%%%%%%%%%%%%%%%%%%%%%%%%%%%%%%%%
	\begin{figure} 
		\centering
		\includegraphics[width=0.90\linewidth]{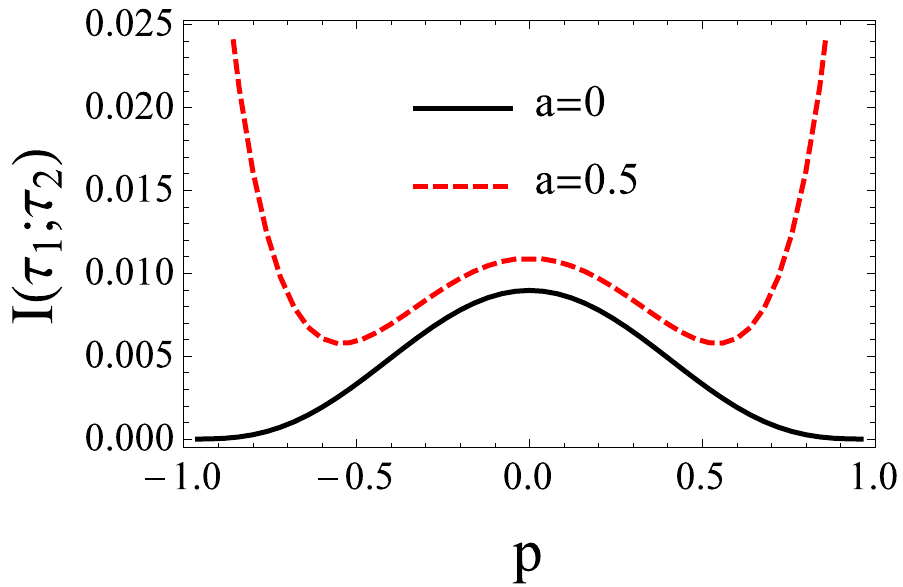} 
		\caption{(Color online). Dependence of the mutual information between distributions of two successive waiting times on $p$ for $a=0$ (black line) and $a=0.5$ (red dashed line).}
		\label{fig:mutinf}
	\end{figure}
	%%%%%%%%%%%%%%%%%%%%%%%%%%%%%%%%%%%%%%%%%%%%%%%%%%%%%%%%%%%%%%%%
	
When only one mechanism generating waiting time correlations is present, the normalized cross-correlation seems to be a proper quantity describing the degree of deviation from the renewal behavior. However, as shown in the previous paragraph, $NCC$ may be equal to 0 even in the case of the nonrenewal dynamics, providing that mechanisms generating both the positive and the negative correlations of waiting times are present. This suggests, that the cross-correlation, although very informative, is not a sufficient indicator of the nonrenewal behavior. Therefore, one can look for an another quantity which is always non-zero when the renewal assumption does not hold. This requirement is met by the mutual information~\cite{cover2006}, which is a popular measure of dependence of two random variables. For the considered case it can be defined in the following way:
\begin{eqnarray} \label{mutinf}
\nonumber I(\tau_1;\tau_2) = \int_0^\infty d \tau_1 \int_0^\infty d \tau_2 w(\tau_1,\tau_2) \log \left[ \frac{w(\tau_1,\tau_2)}{w(\tau_1)w(\tau_2)} \right]. \\
\end{eqnarray}
Dependence of the mutual information on $p$ for two chosen values of $a$ is shown in Fig.~\ref{fig:mutinf}. For $a=0$, when the cascade-like behavior is a sole mechanism generating the nonrenewal statistics, the mutual information decreases with increasing magnitude of $p$ in the same way, as the magnitude of the cross-correlation. For $p=0.5$ no clear relation between the cross-correlation and the mutual information can be found. As a consequence of Jensen's inequality~\cite{cover2006}, the mutual information is always positive when the dynamics is nonrenewal and equals 0 in the opposite case. As a result, the mutual information informs about the degree of deviation from the renewal behavior, but not about the character of correlations. Thus, the mutual information and the normalized cross-correlation may be considered as two complementary quantities characterizing nonrenewal transport statistics.
	
%%%%%%%%%%%%%%%%%%%%%%%%%%%%%%%%%%%%%%%%%%%%%%%%%%%%%%%%%%%%%%%%%%%%%%%%%%%%%%%%%%%%%%%%%%%%%%%%%%%%%%%%%%%%%%%%%%%%%%%%%%%%%%%%%%%%%%%%%%%%%%%
%%%%%%%%%%%%%%%%%%%%%%%%%%%%%%%%%%%%%%%%%%%%%%%%%%%%%%%%%%%%%%%%%%%%%%%%%%%%%%%%%%%%%%%%%%%%%%%%%%%%%%%%%%%%%%%%%%%%%%%%%%%%%%%%%%%%%%%%%%%%%%%
	
\section{\label{sec:conclusions}Conclusions}
Statistics of current fluctuations in the system of two capacitively coupled quantum dots and the Anderson model in the infinite bias limit have been studied applying the Markovian master equation. Using the joint distribution of two successive waiting times between transport events it was demonstrated that both systems exhibit the nonrenewal behavior -- the successive waiting times are correlated. The correlation between waiting times is a result of the electron-electron interaction. Two mechanisms, the telegraphic switching and the cascade-like behavior, were identified to generate the positive and the negative cross-correlation of subsequent waiting times respectively. The former mechanism, occurring in both analyzed systems, is associated with the stochastic switching between different values of tunneling rates, which is a result of their dependence on the charge configuration of the system. The latter mechanism, occurring in the Anderson model, is a result of the nearly-simultaneous occurrence of two electron tunnelings which fill the dot, thus blocking the transport for a certain time. From the formal point of view, the presence of correlations is associated with the form of the jump operator which describes the analyzed tunneling process. Since the waiting time cross-correlation depends on the transport mechanism, it seems to be a useful quantity characterizing the electronic transport in mesoscopic systems. In particular, it can be useful for identifying the cause of the noise enhancement, since its two most common mechanisms -- the telegraphic switching and the dynamical channel blockade -- can be distinguished by the presence or absence of waiting time correlations.
	
The nonrenewal dynamics also has a profound influence on the results provided by two more common approaches to the study of current fluctuations, the full counting statistics and the single waiting time distribution. This influence is associated with the presence of relations between these approaches, which are valid only in the case of the renewal dynamics. When such relations do not hold, it can be inferred that the dynamics is nonrenewal. For example, in this paper, apart from the already reported breaking of identities between cumulants of FCS and WTD, also a nonequivalence of waiting time distributions characterizing the incoming and outgoing tunneling processes was demonstrated. This finding contrasts with the case of the renewal dynamics, when such distributions are equivalent and related to FCS. In consequence, the presence of correlations increases the number of independent measurable quantities, which may be useful for reconstructing the transport mechanism of the system on the basis of current fluctuations statistics. Moreover, the nonrenewal dynamics breaks some relations between the zero-frequency FCS, WTD and the second-order current correlation function $S(\tau)$. This hinders the reconstruction of WTD on the basis of $S(\tau)$, but on the other hand enables to infer the presence of waiting time correlations without direct measurement of WTD, which can extend the analysis of nonrenewal current fluctuations to the systems to which the single electron counting is not applicable.
	
The presented results should be experimentally verifiable using state-of-the-art techniques. Moreover, the paper may give motivation for reanalysis of the already existing experimental data. For example, it would be worthwhile to analyze waiting time correlations in the telegraphic switching system reported by Fricke~\textit{et al.}~\cite{fricke2007} and in the dynamical channel blockade system studied by Gustavsson~\textit{et al.}~\cite{gustavsson2006b}, which should exhibit the nonrenewal and the renewal dynamics, respectively. Theoretical methods used in this paper may also find applications beyond the field of electronic transport, for example in the statistical kinetics of biomolecular systems~\cite{shaevitz2005}, where WTD is already an established tool~\cite{chemla2008, moffitt2010} and Markovian models similar to the ones considered herein are studied~\cite{barato2015}.  
	
\section*{Acknowledgments}
I thank B. R. Bu\l{}ka for the valuable discussion and comments on the manuscript.

\end{document}